\documentclass[prd,11pt,twoside,tightenlines,nofootinbib,showpacs]{revtex4}

\usepackage{graphicx,mathrsfs,amsmath,amssymb,amsthm, amsopn,mathbbol,bm}
\usepackage[dvips]{color}

\newenvironment{Equation*}{\equation}%
{\endequation}
\newenvironment{Eqnarray*}{\eqnarray}%
{\endeqnarray}

\newcommand{\funcd}[2]{\frac{\delta #1}{\delta #2}}
\newcommand{\ii}{\ensuremath{\mathrm{i}}}
\newcommand{\Lag}{\ensuremath{\mathscr{L}}}
\newcommand{\fint}[1]{\ensuremath{\int \frac{\mathrm{d}^4 #1}{(2\pi)^4}}}
\DeclareMathOperator{\im}{Im}
\DeclareMathOperator{\re}{Re}
\DeclareMathOperator{\sign}{sign}
\newcommand{\feynint}[1]{\ensuremath{\int \frac{\mathrm{d}^{d} {#1}}{(2
\pi)^{d}}}}
\DeclareMathOperator{\artanh}{artanh}
\renewcommand{\d}{\ensuremath{\mathrm{d}}}

\begin{document}

\title{Renormalization of Self-consistent Approximation schemes\\ at
  Finite Temperature II:\\
  Applications to the Sunset Diagram} 
\author{Hendrik van Hees, J{\"o}rn Knoll} 
\affiliation{GSI Darmstadt, Planckstra{\ss}e 1, D-64291 Darmstadt} 
\date{January 30, 2002}

\begin{abstract}
  The theoretical concepts for the renormalization of self-consistent Dyson
  resummations, deviced in the first paper of this series, are applied to
  first example cases for the $\phi^4$-theory. Besides the tadpole
  (Hartree) approximation as a novel part the numerical solutions are
  presented which includes the sunset self-energy diagram into the
  self-consistent scheme based on the $\Phi$-derivable approximation or 2PI
  effective action concept.
\end{abstract}


\maketitle

\section{Introduction}
\label{sect-intro}

In the first paper of this series \cite{vHK2001-Ren-I} (in the following
referred as I) we have derived the theoretical concepts for the
renormalization of Dyson equation based resummation schemes at finite
temperature. It could be shown that such truncated self-consistent
approximations can be renormalized with counter term structures solely
defined on the vacuum level of the considered approach, if two conditions
are met: a) the underlying exact theory has to be renormalizable and b) the
approximation scheme has to be based on Baym's $\Phi$-derivable concept
\cite{lw60,baym62,cjt74}, i.e., on a two-particle irreducible (2PI)
effective action principle.  Thus the self-consistent self-energies
$\Sigma$ are generated from a truncated set of 2PI closed diagrams (called
$\Phi$) of the underlying full theory through
\begin{Equation*}
\label{8c}
\Sigma(p) =2 \ii \funcd{\Phi[G]}{G(p)},
\end{Equation*}
where $\Phi[G]$ is a functional of the self-consistent propagator $G(p)$ in
momentum representation (as in I an asterix behind an equation number
implies that the corresponding relations are only valid at the regularized
level, while all other equations are valid also for the renormalized
quantities). To repeat, the main issue is not to render any divergent loops
finite, this has been pursued many times.  The aim is to deploy the counter
term structure such that it is entirely determined at the vacuum level of
the self-consistent scheme. Besides the explicitely visible divergences
this implies to resolve also the nested and overlapping vacuum divergences
hidden in the self-consistent matter parts of the propagator. 

In I we have given the proof that for the self-energy this can be done for
any $\Phi$-derivable approximation, provided the underlying quantum field
theory is renormalizable in the usual sense.  Thereby the counter-term
structure results from closed equations on the vacuum level, implicitly
generating a particular though infinite subset of counter terms. The
complexity of the ensuing equation is similar to that for the self-energy.

Thereby it is of particular importance that the entire counter term
structure is \emph{consistently constructed solely and only within the
  effective action defined by the chosen approximation to $\Phi$}. This
implies that, e.g., the so obtained counter-term scheme and thus the
behavior of the running coupling constant clearly deviates from the true
one at orders beyond those included in $\Phi$. If these rules are not
watched one may face uncompensated divergences as recently encountered in
the detailed work of Braaten and Petitgirard \cite{bp01}. They tackled the
same problem, however using different techniques in terms of a restricted
ansatz for the propagator. Uncompensated singularities arose from the fact
that the used $\beta$-function was not the one pertaining to the
self-consistent scheme but taken from the full $\beta$-function calculated
up to the fifth loop order. Furthermore it is yet not transparent to us,
how in their approach the hidden and nested vacuum subdivergences can be
resolved such that one arrives at a counter term structure solely defined
on the self-consistent vacuum level ($T=0$). In our case the latter leads
to a to be renormalized Bethe-Salpeter equation for the vacuum four-point
function consistent with the chosen approximation level for the
$\Phi$-functional.

In I we also have shown that the generating functional $\Gamma[\varphi,G]$
and thus the thermodynamic potential can be rendered finite with counter
terms solely defined on the vacuum level of the self-consistent scheme.
Again it is crucial to expand the functional around the \emph{solutions of
  the corresponding self-consistent approximation for the vacuum}. It is
not possible to render arbitrary parts of $\Gamma$ finite for themselves
since it is important to use the stationarity of $\Gamma$ for the solution
of the vacuum equations of motion. In particular terms linear in the matter
parts of the propagator are singular and only drop out if the vacuum part
is solved self-consistently!

For self-consistent Dyson resummation schemes it is well known
\cite{baymgrin}, that they may violate symmetry properties such as
crossing symmetry, masslessness of Nambu-Goldstone modes, conservation
laws, etc.  at a level beyond that of one-point functions, i.e., on
the correlator level at orders beyond those included in $\Phi$. We
devote a forthcoming paper \cite{vHK2001-Ren-III} in this series to
the issue of how to extend the scheme such that symmetry violations
are indeed cured.

In this paper we give first numerical solutions for the leading and the
next-to-leading order of the self-consistent Dyson equations for the
$\phi^{4}$-theory, which is defined by the Lagrangian
\begin{equation}
\Lag=\frac{1}{2} (\partial_{\mu} \phi)(\partial^{\mu}
\phi)-\frac{m^{2}}{2} \phi^{2} -\frac{\lambda}{4!} \phi^{4}.
\end{equation}
In the symmetric phase, $\left<\phi\right>=0$, the first diagrams
contributing to $\Phi$ and $\Sigma$ are
\begin{Eqnarray*}
\label{Phi-sunset}
\Phi=&\parbox[][10\unitlength][c]{17\unitlength}{\centerline{\includegraphics{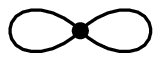}}}
&+ \;\frac{1}{2}\;
\parbox[][10\unitlength][c]{15\unitlength}{\centerline{\includegraphics{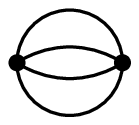}}}
+\cdots, \\
\label{Sigma-sunset}
\Sigma=&\parbox[][15mm][c]{17mm}{\centerline{\includegraphics{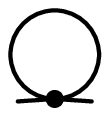}}}
&+\;\;\,\parbox[][10mm][c]{18\unitlength}{\centerline{\includegraphics{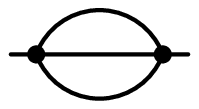}}}
+ \cdots,
\end{Eqnarray*}
where all lines represent self-consistent propagators. The leading
order diagram gives rise to the tadpole (Hartree) approximation for
the self-energy, which, frequently considered in the literature (see
for instance \cite{baymgrin,pesh00}), leads to the standard gap
equation.  Here it is given as the most simple example for our general
renormalization scheme. The next-to-leading order includes the sunset
diagram for the self-energy. While in the tadpole case the self-energy
is real and constant, with the sunset term the self-energy becomes
momentum dependent and complex. Thus the particles acquire a finite
spectral width due to collisions with the surrounding matter. With the
sunset diagram one enters a new stage of sophistication, both, as far
as the counter term structure is concerned, and with respect to the
numerical solution of the resulting self-consistent equations.

First numerical investigations of the self-energy are presented which
result from the two explicitly given diagrams in (\ref{Sigma-sunset}).  We
could improve the numerical accuracy immensely compared to the status given
in \cite{vHK2001} where we restricted ourselves to the computation of the
imaginary parts of the self-energy to avoid problems with renormalization.
The paper is organized as follows: In sect. \ref{sect-eom} we briefly
summarize the results of I as far as we need the formulae for the numerical
calculations. In sect.  \ref{sect-tad} we solve the tadpole approximation
(gap-equation) and in sect. \ref{sect-sun} the next-to-leading order
approximation including the sunset diagram.

Throughout the paper we work in momentum-space representation within
the real-time field theory formalism for the self-energies and propagators.
Thus, if not stated otherwise, self-energies and Green's functions are
contour matrix valued in the sense of the Schwinger-Keldysh real-time
contour \cite{Sch61,kel64,kv97}. There vertices belonging to time arguments
on the time and anti-time ordered branches are labeled by $-$ and $+$
superscripts, respectively. The integration sense is accounted for by
assigning the values $\pm\ii\lambda$ to the bare vertex depending on its
placement on the $\{-\}$ or $\{+\}$ branches of the contour.

\section{The renormalized equations of motion}
\label{sect-eom}

In I we have shown that arbitrary self-consistent approximations given
by the $\Phi$-derivable concept (\ref{8c}) can be renormalized with
counter terms defined on the vacuum level. Thereby the renormalized
self-energy as well as the thermodynamical potential are finite and
consistent to one another.

For the renormalization procedure the self-energy has to be split into
three parts, namely the pure vacuum part (which is of divergence
degree $\delta(\Sigma^{(\text{vac})})=2$), the part $\Sigma^{(0)}$
with divergence degree $\delta(\Sigma^{(0)})=0$, containing explicit
and hidden vacuum sub-divergences, and the rest $\Sigma^{(\text{r})}$
with $\delta(\Sigma^{(\text{r})})=-2$, which contains only explicit
vacuum sub-divergences with corresponding counterterms
\begin{equation}
\label{1}
\Sigma=\Sigma^{(\text{vac})} + \Sigma^{(0)} + \Sigma^{(\text{r})}.
\end{equation}
The self-consistent Green's function $G$ follows from the Dyson equation in
contour matrix form
\begin{equation}
\label{Dyson}
\Delta^{-1} G=1  +\Sigma\; G.
\end{equation}
Here $\Delta$ is the free Green's function on the contour. In view of
(\ref{1}) the self-energy within the Green's function needs two
subtractions to render its loops finite:
\begin{equation}
\label{2}
\begin{split}
  \ii G &= \ii G^{(\text{vac})}+\underbrace{\ii G^{(\text{vac})}
    \Sigma^{(0,\text{div})} G^{(\text{vac})} + \ii G^{(\text{r})}}_{\ii
    G^{(\text{matter})}}
  =\parbox[][5mm][c]{15mm}{\includegraphics{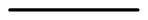}} +
    \parbox[][10mm][c]{20mm}{\includegraphics[width=20mm]{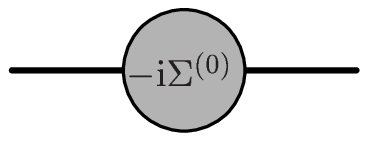}} +
    \parbox[][6mm][c]{15mm}{\includegraphics{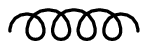}}.
\end{split}
\end{equation}
The subtraction only affects contour diagonal pieces of the
propagator, since loops containing mixed contour vertices are finite
for themselves. Thus
\begin{equation}
\label{GvacS0}
\ii G^{(\text{vac})}(p)=
\left(\begin{array}{cc}
\ii G^{--(\text{vac})}(p)&0\\ 0&\ii G^{++(\text{vac})}(p)
\end{array}\right),
\quad
\ii \Sigma^{(\text{0;div})}(p)=
\left(\begin{array}{cc}
\ii \Sigma^{--(0)}(p)&0\\ 0&\ii \Sigma^{++(0)}(p)
\end{array}\right),
\end{equation}
and likewise $\Sigma^{\text{(vac)}}$ are diagonal. While $G^{(\text{vac})}$
and $\Sigma^{(\text{vac})}$ are given by the (anti-)time ordered
self-consistent expressions given by the $\Phi$-derivable scheme on the
vacuum level, $\Sigma^{(0)}$ accounts for the self-energy parts linear in
$G^{\text{(matter)}}$. The three r.h.s. terms of (\ref{2}) are of
divergence degree $-2$, $-4$, and $-6$, respectively.  It is important to
notice that only the full self-consistent propagator and self-energy, $G$
and $\Sigma$, obey the equilibrium conditions (KMS) among their four
contour components, cf.  Eqs. (A.14) to (A.18) in I. The components of
subtracted pieces like $G^{\text{(matter)}}$, $\Sigma^{(0)}$ or
$\Sigma^{\text{(r)}}$ obey no direct equilibrium relations as they result
from differences of finite temperature with vacuum terms. Here we see an
advantage in the real-time formalism which naturally permits this
decomposition into vacuum and matter pieces of all dynamical quantities.

The contour diagonal parts of $\Sigma^{\text{(0)}}$ and
$\Sigma^{(\text{r})}$ contribute only at finite temperature. Due to
Lorentz invariance the vacuum self-energy and the Green's function are
functions of $s=p^{2}$. The vacuum self-energy is renormalized
according to the rules of the on-shell renormalization scheme, i.e.,
\begin{equation}
\label{16}
\Sigma^{(\text{vac})}(s=m^{2})=0, \quad
\partial_{s}\Sigma^{(\text{vac})}(s=m^{2})=0, 
\end{equation}
which defines $m$ to be the physical mass of the particles and normalizes
the wave functions such that the residuum of the Green's function at
$s=m^{2}$ is equal to $1$. As we shall see in the following the
renormalization procedure used to calculate the vacuum parts, and
subsequently needed for the temperature dependent pieces of the
self-energy, allows to chose any renormalization scheme.  Especially it is
possible to use ``mass independent'' minimal subtraction schemes (MS
\cite{hooft73} or $\overline{\text{MS}}$ \cite{bard78} of dimensional
regularization) so that the renormalization program shown in I is
applicable also for the massless case. For the renormalization of the
self-energy and the 2PI-functional $\Gamma[G]$ it is crucial to split all
quantities into vacuum ($T=0$) and finite temperature parts by the
procedure given above.

As shown in I this procedure allows to define the generating functional in
its dependence on the in-matter parts of the propagator with the
self-consistent solution given by the stationary point.  Simultaneously
this leads to the renormalized thermodynamical potential. All divergences
indeed compensate if and only if the vacuum level is solved and
renormalized entirely within the same approximation level as used for the
finite temperature parts. This fact together with the \emph{stationarity
  conditions of the generating functional at $T=0$} for the vacuum
equations of motion avoids the singularity problems encountered in
\cite{bp01}.

In $\Sigma^{(0)}$ the in-matter part of the propagator is involved in
logarithmically divergent loops which imply hidden divergences. As
shown in I these can be resolved with the following result
\begin{equation}
\label{5}
  \Sigma^{(0)}(p) =\fint{l} \left \{ [\Gamma^{(4)}(l,p) -
  \Gamma^{(4,\text{vac})}(l,0)] G^{(\text{matter})}(l) + \Lambda(0,l)
  G^{(\text{r})}(l) \right \}.
\end{equation}
Here in consistency with the Dyson equation the four-point function
$\Gamma^{(4)}(p,q)$ is generated also from $\Phi$ through
\begin{equation}
\label{4}
\Gamma^{(4)}(p,q)= 2 \left( \frac{\delta^2{\Phi[G]}}{\delta
  G(p) \delta G(q)}\right)_{T \rightarrow 0^{+}}^{\text{(ren)}}.
\end{equation}
The counter term $\Gamma^{(4,\text{vac})}$ contains only the contour
diagonal parts (all vertices placed on one side of the real-time contour)
of $\Gamma^{(4)}$. This function is 2PI in the channel $p \rightarrow q$,
i.e., one cannot disconnect the diagrams by cutting two lines and
separating the lines carrying the momenta $p$ and $q$.  It contains only
vacuum pieces and can straight forwardly be renormalized with
$\Gamma^{\text(4,\text{vac})}(0,0)=\pm\lambda/2$ (denoted by the
superscript (ren)).  At the same time this four-point function defines the
kernel for the vacuum $s$-channel Bethe-Salpeter equation defining the 2PR
four-point function $\Lambda(p,q)$. In (\ref{5}) we only need the half
sided $\Lambda(0,p)$, which, as one of the crucial points of the
Bogoliubov-Parasiuk-Hepp-Zimmermann (BPHZ) renormalization procedure
\cite{bp57,Zim69}, is given by the finite equation
\begin{equation}
\label{3}
\begin{split}
  \Lambda(0,p) = &\pm\frac{\lambda}{2}+\Gamma^{(4,\text{vac})}(0,p) +
  \ii \fint{l} \Lambda(0,l)[G(l)]^{2} [\Gamma^{(4,\text{vac})}(l,p)
  -\Gamma^{(4,\text{vac})}(l,0)],
\end{split}
\end{equation}
again involving only contour diagonal terms.  Here we inserted the
renormalization condition $\Lambda(0,0)=\pm\frac{\lambda}{2}$.

The expression (\ref{5}) for $\Sigma^{(0)}$ void of the $\Lambda(0,l)$
part would correspond to a naive subtraction, which indeed is finite,
however it implies $T$ dependent counter terms. As explained in I,
only the entire expression (\ref{5}) guarantees a counterterm
structure solely defined on the self-consistent vacuum level.\\

\begin{footnotesize}
\noindent{\bf The perturbative view}

The above procedure defines the self-consistent renormalization scheme
in terms of the corresponding self-consistent vacuum propagator
$G^{(\text{vac})}$, the vacuum four-point function $\Lambda(0,l)$, and
the self-consistent in-matter pieces of the propagator
$G^{(\text{matter})}$ or $G^{(\text{r})}$. As an illustration we
briefly quote the corresponding perturbative (1PI) view of the
scheme, expanding the full self-energy in terms of the vacuum
propagator in contour matrix notation satisfying the equilibrium
relations (KMS) at the given temperature $T$
\begin{equation}
\label{GvacT}
\begin{split}
\ii G^{(\text{vac};T)}(p)&=
\parbox{45mm}{\centerline{
$\left(\begin{array}{cc}
\ii G^{--(\text{vac})}(p)&0\\ 0&\ii G^{++(\text{vac})}(p)
\end{array}\right)$}}
+
\rho^{(\text{vac})}(|p_0|,\vec{p})
\left(\begin{array}{cc}
n(p_0)& \Theta(-p_0)+ n(p_0)\\ \Theta(p_0)+ n(p_0)&n(p_0)
\end{array}\right), \\
&=\parbox{45mm}{\centerline{ \includegraphics{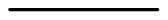}}} +
\parbox{45mm}{\centerline{\includegraphics{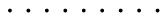}}}
\end{split}
\end{equation}
with the vacuum spectral function and thermal Bose-Einstein weight
defined by
\begin{equation}
\begin{split}
\rho^{(\text{vac})}(p)&=-2\im G^{(\text{vac})}_R(p)
=-2\sign(p_0)\im G^{--(\text{vac})}\\
n(p_0)&=\frac{1}{\exp(\beta|p_0|)-1}.
\end{split}
\end{equation}
As discussed in I, loops involving the $\Theta$-functions or the
thermal weights summarized by the dotted line are always finite.

For the self-consistent scheme of tadpole and sunset the following
types of 1PI diagrams are resummed to the full self-consistent
self-energy:
\begin{Equation*}
\label{Sigma-diag}
\Sigma=\underbrace{\parbox{11mm}{\includegraphics{Tadpole.eps}}+
 \parbox{19mm}{\includegraphics{Sunset.eps}}}_{\mbox{$\Sigma^{(\text{vac})}$}}+
\underbrace{\underbrace{\cdots +
    \raisebox{4mm}{\parbox{19mm}{\includegraphics{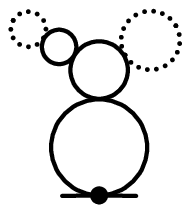}}} +
    \cdots +
    \raisebox{1.3mm}{\parbox{22mm}{\includegraphics{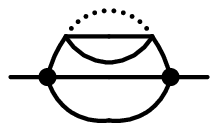}}
        \vphantom{+\raisebox{3.087mm}{%
            \parbox{22mm}{\includegraphics{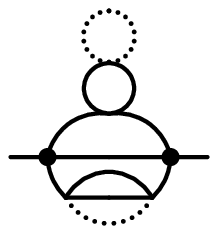}}}}}}
               _{\mbox{$\Sigma^{(0)}$}}
  +
  \underbrace{\cdots+\raisebox{3.87mm}{\parbox{22mm}{\includegraphics{Mixed12.eps}}}
  + \cdots}_{\mbox{$\Sigma^{(\text{r})}$}}}
_{\mbox{$\Sigma^{(\text{matter})}$}}. 
\end{Equation*}
Here the diagrams for $\Sigma^{(\text{matter})}$ are generated by
iterative insertions of the diagrams of $\Sigma^{(\text{vac})}$,
replacing one or more plain lines by dotted ones with the constraint
that due to the self-consistence at the vacuum level {\em all pure vacuum
self-energy insertions are to be excluded}. In
$\Sigma^{(\text{matter})}$ vacuum subdiagrams (plain lines) with 4
external lines are logarithmically divergent and need to be
renormalized. Their sum defines the renormalized four point function
$\Lambda$. If the latter include the external points (full dots) they
contribute to $\Sigma^{(0)}$, defined in (\ref{1}). All other terms,
like the last term, contribute to $\Sigma^{(\text{r})}$.

\end{footnotesize}

\section{The tadpole approximation}
\label{sect-tad}

\begin{figure}
\centering{\includegraphics[width=0.6 \textwidth]{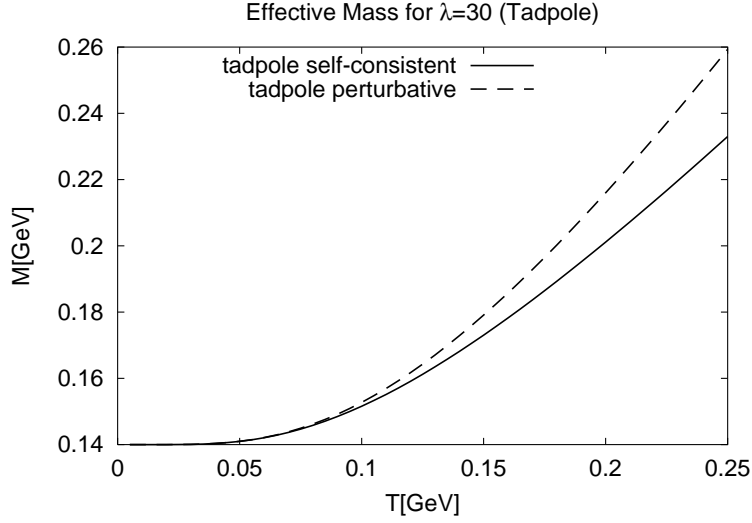}}
\caption{The solution of the gap equation (\ref{13}) compared to the
  naive, i.e., non-resummed one-loop perturbative solution.
  Both calculations are renormalized in the vacuum in the same way
    according to the on-shell conditions (\ref{16}) with
  $m=0.14\text{ GeV}$}.
\label{fig-1}
\end{figure}

The tadpole approximation is given by
\begin{Eqnarray*}
\label{6}
\ii \Phi[G] =&
\parbox{16\unitlength}{\centerline{\includegraphics{eight.eps}}}
&= \frac{\ii \lambda}{8} \sum_{j=\pm} \sigma_{jj} \left [\feynint{l}
  G^{jj}(l) \right]^2\\
\label{7}
-\ii \Sigma^{jj} = 2 \ii \funcd{\Phi[G]}{G_{jj}}=&
\raisebox{3mm}{\parbox{11mm}{\centerline{\includegraphics{Tadpole.eps}}}} &=
\frac{\lambda}{2} \feynint{l} G^{jj}(l) = \Theta \sigma^{jj}.
\quad j\in \{-,+\}
\end{Eqnarray*}
for the un-renormalized $\Phi$-functional and self-energy. Here
$\sigma=\text{diag}(1,-1)$ accounts for the integration sense on the
contour. The self-energy is diagonal in the contour indices. Since the
tadpole self-energy is real and constant the vacuum part vanishes
\begin{equation}
\Sigma^{(\text{vac})} = 0
\end{equation}
in view of the physical renormalization condition (\ref{16}). To find
the renormalized equations of motion at finite temperature we need the
renormalized four-point function $\Lambda$ according to (\ref{5}). Due
to (\ref{4}) we have a constant Bethe-Salpeter kernel
\begin{equation}
\label{9}
\Gamma^{(\text{4;vac})} = \frac{\lambda}{2}
\end{equation}
and thus $\Lambda$ is also constant and determined from its
renormalization condition (\ref{3})
\begin{equation}
\label{10}
\Lambda^{(\text{vac})}=\frac{\lambda}{2}.
\end{equation}
In the following it is sufficient to just consider the $\{--\}$-propagator.
Writing $M^{2}=m^{2}+\Theta$ from (\ref{Dyson}) the self-consistent
$\{--\}$-propagator is given by
\begin{equation}
\label{11}
\begin{split}
G^{--}(p) &= \frac{1}{p^2-M^2+\ii \eta} - 2\ii \pi n_T(p_{0})
\delta_{\eta}(p^2-M^2) \\
\text{with }\quad & \delta_{\eta}(x)=\frac{1}{\pi} \im
\frac{1}{x-\ii \eta}=\frac{1}{\pi} \frac{\eta}{x^2+\eta^2}
\end{split}
\end{equation}
and the Bose-Einstein function 
\begin{equation}
n_T(p_{0})=\frac{1}{e^{|p_{0}|/T}-1}.
\end{equation}
Since the tadpole self-energy is constant it is identical with
$\Sigma^{(0)}$ and due to (\ref{2}) we thus have
\begin{equation}
\label{12}
\begin{split}
  G^{--(\text{r})}(p) &=
  G^{--}(p)-G^{--(\text{vac})}(p)-\Theta[G^{--(\text{vac})}(p)]^2 \\ &=
  \frac{\Theta^2}{(m^{2}-p^{2}-\ii \eta)^2(M^{2}-p^{2} - \ii \eta)}- 2 \pi\ii
  n_T(p_{0}) \delta_{\eta}(p^2-M^2).
\end{split}
\end{equation}
This analytically given result explicitly illustrates that the subtracted
part $G^{--(\text{r})}$ of the propagator no longer obeys the finite
temperature KMS and retarded relations (Eqs. (A.14)-(A.18) in I). The
renormalized effective mass follows from (\ref{5}) and (\ref{10}) with
standard integrals from vacuum-perturbation theory (see for instance
\cite{ram89}):
\begin{equation}
\label{13}
M^{2}= m^{2} + \Theta =m^{2}+ \frac{\lambda}{32 \pi^{2}} \left [M^{2}
  \ln \left  (\frac{M^{2}}{m^{2}} \right) - M^{2}+m^{2} \right] + 
  \frac{\lambda}{4 \pi^{2}} \int_{0}^{\infty} \d L L^{2}
  \frac{n_T(\sqrt{L^{2}+M^{2}})}{\sqrt{L^{2}+M^{2}}}.
\end{equation}
The solution of this gap-equation for $m=0.14\,\text{GeV}$ and
$\lambda=30$ is shown in Fig. \ref{fig-1}.
The self-consistent treatment reduces the mass gap relative to the
perturbative result due to the larger mass $M$ entering the
self-consistent loop.

The gap equation (\ref{13}) becomes singular for $m \rightarrow 0$ due to
the chosen on-shell renormalization condition (\ref{16}). For the sake of
completeness we give a brief summary about the treatment of the case of
vanishing renormalized vacuum mass in the following.

For $m=0$ the renormalization description has to be changed to a so-called
``mass independent'' renormalization scheme. This concept is most easily
established in the dimensional regularization procedure by the so called
minimal subtraction (MS) or modified minimal subtraction
($\overline{\text{MS}}$) schemes where for all renormalization parts, i.e.,
in our case of $\phi^{4}$-theory the proper $n$-point vertex functions with
$n \leq 4$, only the singular terms in $\epsilon=(4-d)/2$ are canceled.

In the present case of the tadpole approximation where the self-energies
are momentum-independent this can be done analytically. Indeed for $m=0$ we
have $\Theta=M^{2}$ and the divergent part of the diagram reads
\begin{Equation*}
\label{14n}
\Theta_{\infty}=\frac{\ii \lambda \mu^{2 \epsilon}}{2}
\feynint{l} \frac{1}{l^2-M^{2}+\ii \eta} = -\frac{\lambda M^{2}}{32
  \pi^{2}} \left[\frac{1}{\epsilon} + 1 -\gamma - \ln \left (\frac{M^{2}}{4
      \pi \mu^{2}} \right) + O(\epsilon) \right] 
\end{Equation*}
in dimensional regularization. Since now the renormalized vacuum propagator
is the free massless propagator, i.e.,
\begin{equation}
\label{15n}
G_{\text{vac}}(l)=\frac{1}{l^2+\ii \eta}
\end{equation}
an infrared regulating mass scale $\mu'$ has to be introduced in order to
calculate both, the mass and the four-point vertex
\begin{Equation*}
\label{16n}
\begin{split}
\Theta_{\text{vac}} &= \frac{\ii \lambda \mu^{2 \epsilon}}{2}
\feynint{l} \frac{1}{l^2-\mu'{}^{2}+\ii \eta} = -\frac{\lambda \mu'{}^{2}}{32
  \pi^{2}} \left[\frac{1}{\epsilon} + 1 -\gamma - \ln \left (\frac{\mu'{}^{2}}{4
      \pi \mu^{2}} \right) + O(\epsilon) \right],\\
\Gamma_{\text{vac}}^{(4)}(0) & = \frac{\ii \lambda \mu^{2 \epsilon}}{2}
\feynint{l} \frac{1}{(l^2-\mu'{}^{2}+\ii \eta)^2} = -\frac{\lambda}{32
  \pi^{2}} \left[\frac{1}{\epsilon} -\gamma - \ln \left (\frac{\mu'{}^{2}}{4
      \pi \mu^{2}} \right) + O(\epsilon) \right]
\end{split}
\end{Equation*}
with the $\overline{\text{MS}}$-counter terms
\begin{Equation*}
\label{17n}
\begin{split}
  \delta m^{2} &= \lim_{\mu' \rightarrow 0} \frac{\lambda \mu'{}^{2}}{32
    \pi^{2}} \left[\frac{1}{\epsilon} -\gamma \right]=0, \\
\delta \lambda &= \frac{\lambda}{32
    \pi^{2}} \left[\frac{1}{\epsilon} -\gamma \right],
\end{split}
\end{Equation*}
which are IR-finite due to fact that in dimensional regularization the mass
counter term in the $\overline{\text{MS}}$ scheme is proportional to the
infrared-regulator mass $\mu'{}^{2}$.  For $m=0$ the renormalized gap
equation finally takes the form
\begin{equation}
\label{18n}
M^2=-\frac{\lambda M^{2}}{32 \pi^{2}} \left[1-\ln \left(\frac{M^{2}}{4 \pi
      \mu^{2}} \right)\right]  + \frac{\lambda}{4 \pi^{2}}
      \int_{0}^{\infty} \d L L^{2}
      \frac{n_T(\sqrt{L^{2}+M^{2}})}{\sqrt{L^{2}+M^{2}}}
\end{equation}
implying that perturbatively
\begin{equation}
\label{19n}
M_{\text{pert}}^2= \frac{\lambda}{4 \pi^{2}} \int_{0}^{\infty} \d L L
n_T(L) = \frac{\lambda}{24} T^{2}
\end{equation}
which follows from (\ref{18n}) by letting $M \rightarrow 0$ on the r.h.s.

\section{The sunset approximation}
\label{sect-sun}

In this section we calculate the self-consistent self-energy for the
next-to-leading order approximation of the $\Phi$-functional
numerically. First we have to solve the renormalized Dyson equation
for the vacuum and the Bethe-Salpeter ladder equation for
$\Lambda^{(\text{vac})}$, which is needed as input for the temperature
dependent calculation according to (\ref{5}).

The $\Phi$-functional is given diagrammatically by
\begin{Equation*}
\label{14}
\Phi=\parbox[c][10\unitlength][c]{16\unitlength}{\includegraphics{eight.eps}} +
\frac{1}{2}
\parbox[c][15\unitlength][c]{15\unitlength}{\includegraphics{baseball.eps}},
\end{Equation*}
while the corresponding self-energy and BS-kernel become
\begin{Eqnarray*}
\label{15}
\Sigma=&\raisebox{4.5mm}{\parbox{7mm}{\centerline{\includegraphics{Tadpole.eps}}}}&
+\parbox{20mm}{\centerline{\includegraphics{Sunset.eps}}}\\
\label{15a}
\Gamma^{(4)}=&\parbox{7mm}{\centerline{\includegraphics{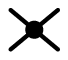}}}&+\parbox{20mm}{\centerline{\includegraphics{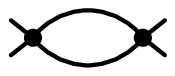}}}.
\end{Eqnarray*}

For the renormalization parts it is sufficient to restrict all
considerations to one real-time contour branch. For definiteness we choose
the time-ordered branch, i.e., in the following subsections all contour
two-point functions denote $\{--\}$-quantities rather than matrices.  The
physical renormalization scheme (\ref{16}) implies that the tadpole part of
the self-energy is already subtracted such that in the vacuum case we only
need to solve for the self-consistent sunset self-energy.

The general strategy will be to combine the BPHZ-renormalization scheme
with dimensional regularization and use the spectral representation for the
Green's functions (Lehmann representation)
\begin{equation}
\label{18}
G^{(\text{vac})}(p^{2})=\int_{0}^{\infty} \frac{\d (m^2)}{\pi} \frac{\im
  G^{(\text{vac})}(m^2)}{m^2-p^2-\ii \eta}.
\end{equation}
All quantities like the renormalized vacuum self-energy
$\Sigma^{(\text{vac})}$, the BS-kernel $\Gamma^{(4)}$ and the four-point
function $\Lambda$ are then to be expressed through the vacuum spectral
function $\rho=-2\im G^{(\text{vac})}$ and corresponding time-ordered
kernel functions $K_i$. The latter express the analytic structure entirely
in terms of free particle properties, however with varying masses. They can
then be renormalized by dimensional regularization.

\subsection{The two-propagator loop}

A central quantity is the simple loop-function contained in (\ref{15})
and (\ref{15a})
\begin{equation}
\label{18a}
L^{(\text{reg})}(q^{2})=\ii \, \raisebox{0.2mm}{\parbox{17mm}{\centerline{\includegraphics{L-diag.eps}}}}=\ii \feynint{l} G^{(\text{vac})}[(l+q)^{2}]
G^{(\text{vac})}(l^{2}). 
\end{equation}
This four-point function is logarithmitically divergent and to be
renormalized with the condition
\begin{equation}
\label{19}
L^{(\text{ren})}(0)=0.
\end{equation}
With help of the Lehmann representation (\ref{18}) it can be expressed
as
\begin{equation}
\label{20}
L^{(\text{ren})}(q^2)=\int_{0}^{\infty} \frac{\d m_{1}^{2}}{\pi}
\int_{0}^{\infty} 
\frac{\d m_{2}^{2}}{\pi}
K_{1}^{(\text{ren})}(q^{2},m_{1}^{2},m_{2}^{2}) \im
G^{(\text{vac})}(m_{1}^{2}) \, \im G^{(\text{vac})}(m_{2}^{2})
\end{equation}
with the renormalized kernel defined through
\begin{equation}
\label{21}
\begin{split}
K_{1}^{(\text{ren})}(q^{2},m_{1}^{2},m_{2}^{2}) &=
K_{1}^{(\text{reg})}(q^{2},m_{1}^{2},m_{2}^{2}) -
K_{1}^{(\text{reg})}(0,m_{1}^{2},m_{2}^{2}), \\
K_{1}^{(\text{reg})}(q^{2},m_{1}^{2},m_{2}^{2}) &= \ii \feynint{l}
\frac{\mu^{2 \epsilon}}{(m_{1}^2-l^{2}-\ii \eta)[m_{2}^2-(l+p)^{2}-\ii
  \eta]}. 
\end{split}
\end{equation}
Here the standard notation for dimensionally regularized quantities is
used, where $d=4-2 \epsilon$ is the space-time dimension, and $\mu$
denotes the \emph{renormalization scale}.

After a Feynman parameterization the integral (\ref{21}) can be
expressed in terms of standard formulae of dimensional regularization
(see, e.g., \cite{ram89}) which leads to the result
\begin{equation}
\label{23}
\begin{split}
K_{1}^{(\text{reg})}(s,m_1^2,m_2^2)  = 
& \frac{1}{16 \pi^2 s} \Bigg \{ - \left ( \frac{1}{\epsilon}+2 -
  \gamma \right) s + \\ 
& \quad + \lambda(s,m_{1},m_{2}) \left [ \artanh \left
    (\frac{s + m_{1}^{2}-m_{2}^{2}}{\lambda(s,m_{1}^{2},m_{2}^{2})}
  \right) +  \artanh \left
    (\frac{s-m_{1}^{2}+m_{2}^{2}}{\lambda(s,m_{1}^{2},m_{2}^{2})}
  \right) \right] + \\
& \quad + (m_{1}^{2}-m_{2}^{2}) \ln \left(\frac{m_{1}}{m_{2}} \right) + s \ln
\left (\frac{m_{1} m_{2}}{4 \pi \mu^2} \right) \Bigg \}. 
\end{split}
\end{equation}
with the K\"all\'en function
\begin{equation}
\label{22}
\lambda(s,m_{1}^2,m_{2}^2)=\sqrt{s^2+m_{1}^4+m_{2}^4-2s m_{1}^2+-2s
  m_{2}^2 -2m_1^2m_2^2},
\end{equation}
which is completely symmetric in its 3 arguments and the branch is
determined by the $s+\ii\eta$ prescription.  Here and in the following
we neglect contributions of order $O(\epsilon)$. For the
renormalization of this kernel we also need its value at $s=0$ given
by
\begin{equation}
\label{24}
\begin{split}
  K_{1}^{(\text{reg})}(0,m_{1},m_{2})= \frac{1}{16 (m_{1}^2-m_{2}^2) \pi^2}
  \Bigg [ & -\frac{m_{1}^{2}-m_{2}^2}{\epsilon} -1 +\gamma \\ & + 
  m_{1}^{2} \ln \left ( \frac{m_{1}^{2}}{4 \pi \mu^2} \right)  -
  m_{2}^{2} \ln \left(\frac{m_{2}^{2}}{4 \pi \mu^2} \right) \Bigg ].
\end{split}
\end{equation}
According to (\ref{21}) we thus find
\begin{equation}
\label{25}
\begin{split}
K_{1}^{(\text{ren})}(s,m_{1},m_{2}) & =  [K_{1}^{(\text{reg})}(s)-
K_{1}^{(\text{reg})}(0)]|_{
\epsilon \rightarrow 0} \\ & = \frac{1}{16 \pi^{2} s} \Bigg
 \{- s + \frac{(m_{1}^{2}-m_{2}^{2})^{2}-s (m_{1}^{2}+m_{2}^{2})}{m_{1}^{2}-m_{2}^{2}} 
\ln \left ( \frac{m_{1}}{m_{2}} \right) \\ 
& \quad + \lambda(s,m_{1},m_{2}) \left [ \artanh \left
    (\frac{s + m_{1}^{2}-m_{2}^{2}}{\lambda(s,m_{1}^{2},m_{2}^{2})}
  \right) +  \artanh \left
    (\frac{s-m_{1}^{2}+m_{2}^{2}}{\lambda(s,m_{1}^{2},m_{2}^{2})}
  \right) \right] \Bigg \}. 
\end{split}
\end{equation}
\begin{figure}
\begin{minipage}{0.49\textwidth}
\centering{\includegraphics[width=\textwidth]{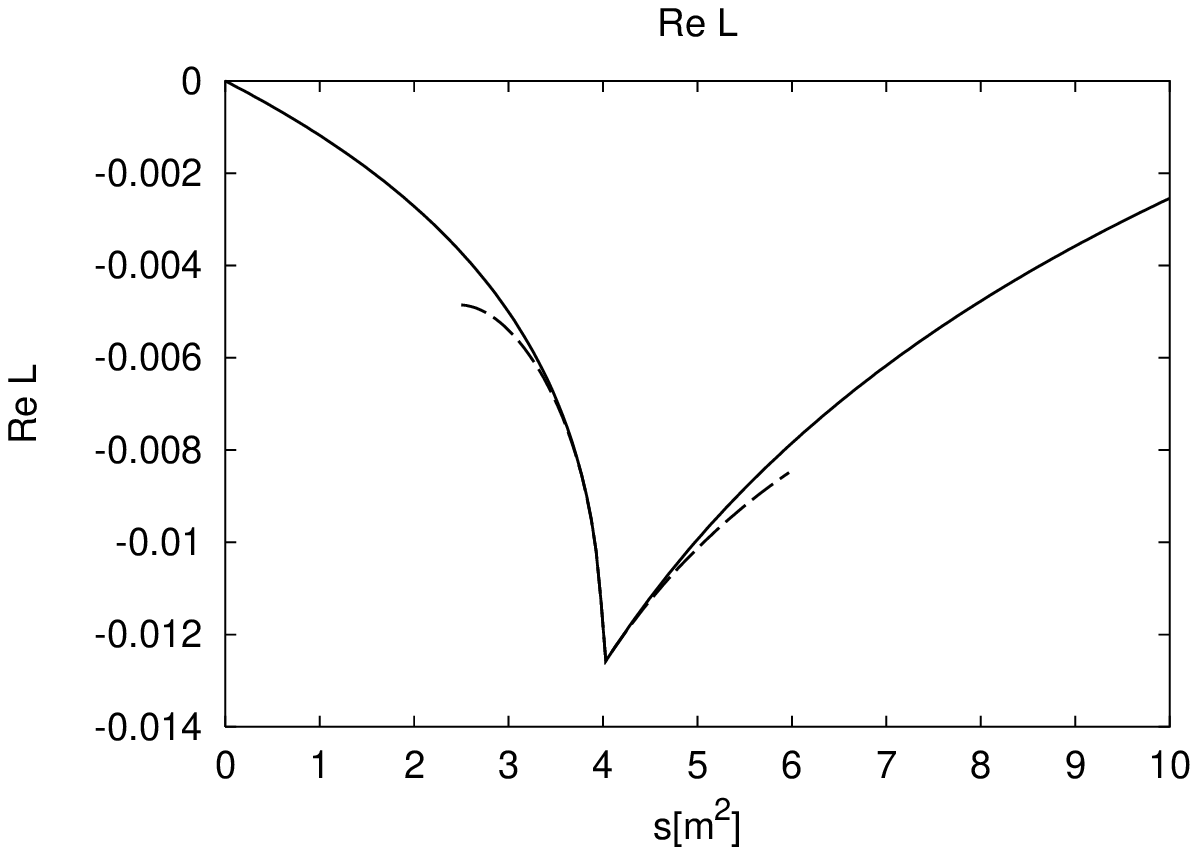}}
\end{minipage}
\begin{minipage}{0.49\textwidth}
\hspace{2mm}
\centering{\includegraphics[width=\textwidth]{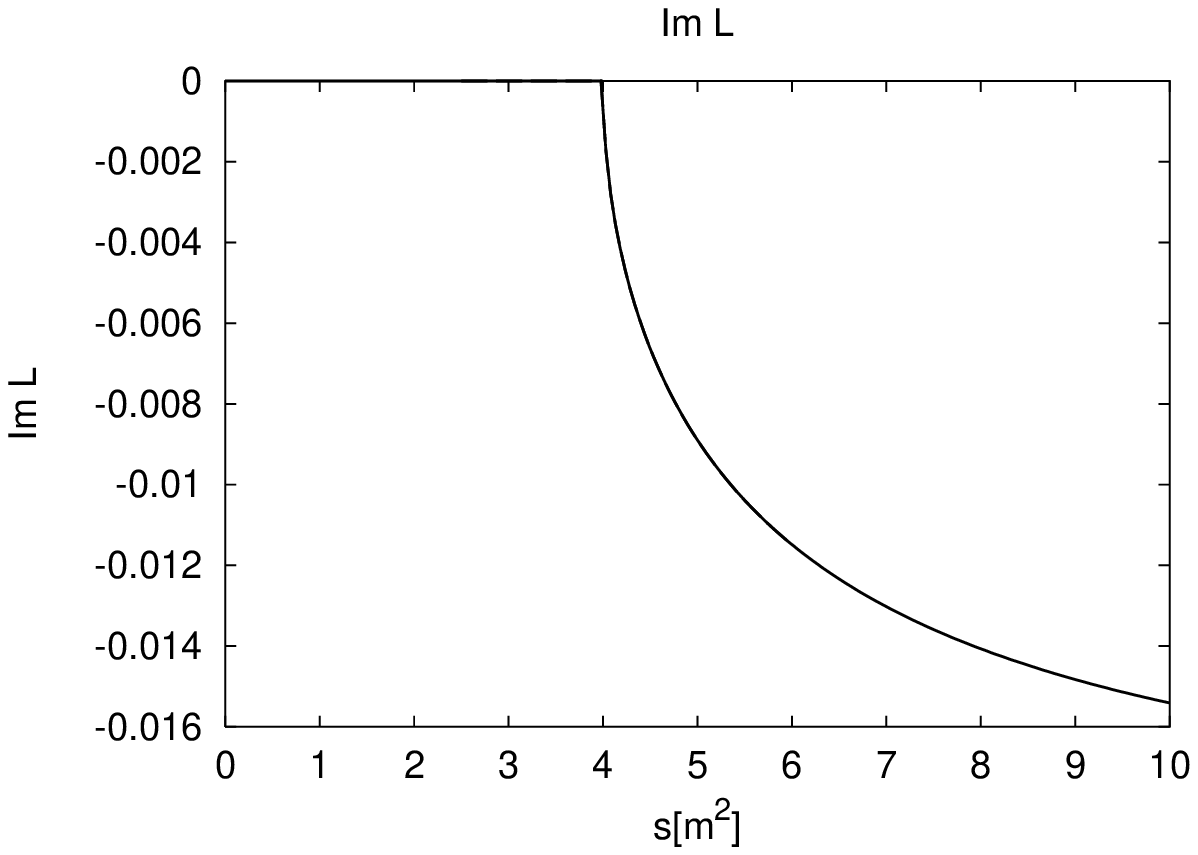}}
\end{minipage}
\caption{Real (left) and imaginary part (right) of the loop
  function $L$. For the real part the dashed line shows the
  approximate behavior (\ref{L-asymp}) around the threshold.}
\label{Fig-L}
\end{figure}
The result for the loop $L(s)$ with $s=q^2$ is given in Fig.
\ref{Fig-L}. The imaginary part is essentially determined by the
two-body phase-space which opens at $s=4m^2$.  To discuss the
threshold singularity of $L$ it is sufficient to study the
perturbative result obtained from the pole term of $G^{(\text{vac})}$
which is analytically given as
\begin{equation}
\label{a.1}
\begin{split}
  L^{(\text{pert})}(s,m^2) &= K_1^{(\text{ren})}(s,m^2,m^2)=\frac{1}{8
    \pi^2} \left [\sqrt{\frac{s-4m^{2}}{s}} \artanh \left (
      \sqrt{\frac{s}{s-4m^{2}}} \right) -1 \right] \\
&=\frac{1}{8 \pi^{2}} \left [\sqrt{\frac{s-4m^{2}}{s}}  \left ( \ln
    \frac{\sqrt{s}+\sqrt{s-4 m^{2}}}{2m} - \frac{\ii \pi}{2} \right) -1
  \right].
\end{split}
\end{equation}
This function is analytic and real for real $s<0$.  The analytic
continuation to other values of $s$ is given by taking the principle
branch of the square root and the $\artanh$ functions for $s+\ii \eta$
with a small $\eta>0$. Close to threshold $s\approx 4 m^2$ one
obtains the following approximate forms
\begin{equation}
\label{L-asymp}
8\pi^2 L^{(\text{pert})}(s,m^2) = \left\{
\mbox{$\begin{array}{ccccl}\displaystyle
-\frac{\pi}{2}\sqrt{\frac{4m^2-s}{s}}
&-&\displaystyle\frac{4m^2}{s}+O\Big(\Big(\frac{s-4m^2}{s}\Big)^2\Big)
\;&\text{\normalsize for}\;&
\text{\normalsize $0<4m^2-s\ll s$},
\\[2mm]
\displaystyle\frac{\ii\pi}{2}\sqrt{\frac{s-4m^2}{s}}
&-&\displaystyle\frac{4m^2}{s}+O\Big(\Big(\frac{s-4m^2}{s}\Big)^2\Big)
\;&\text{\normalsize for}\;&
\text{\normalsize $0<s-4m^2\ll s$}.
\end{array}$}\right.
\end{equation}
Around the threshold the square root term causes the singular behavior
changing from real to imaginary values, while the remaining log-term
is entirely real. At large $s$ the real part of $L$ behaves
logarithmically, whereas the imaginary part goes to a constant.

\subsection{The vacuum self-energy}

The un-renormalized expression of the two-loop self-energy reads
\begin{Equation*}
\label{17}
\Sigma^{(\text{vac})}(p^2)=-\frac{\lambda^2}{6} \feynint{l_1}
\feynint{l_2} G^{(\text{vac})}(l_1^2) G^{(\text{vac})}[(l_1+l_2+p)^2]
G^{(\text{vac})}(l_2^2).
\end{Equation*}
With help of (\ref{20}) the sunset self-energy with all
sub-divergences subtracted becomes
\begin{equation}
\label{26}
\bar{\Sigma}^{(\text{vac})}(p)=\frac{\ii \lambda^2}{6} \feynint{l_2}
L[(l_2+p)^2] G^{(\text{vac})}(l_2^2).
\end{equation}
Now we apply the spectral representation for (\ref{18a}) with one
subtraction determined by (\ref{19})
\begin{equation}
\label{27}
L(q^2)=\int_{4 m^2}^{\infty} \frac{\d (m_{3})^{2}}{\pi} \im
  L(m_{3}^{2}+\ii \eta) \left (\frac{1}{m_{3}^{2}-q^2-\ii \eta} -
    \frac{1}{m_{3}^{2}-\ii \eta} \right ). 
\end{equation}
Together with the Lehman representation of the propagator (\ref{18})
this leads to the renormalized vacuum self-energy
\begin{equation}
\label{28}
\Sigma^{(\text{vac})}(s)=\int_{4 m^2} \frac{\d (m_{3}^{2})}{\pi}
\int_{9m^2} \frac{\d
  (m_{4}^{2})}{\pi} K_{2}^{(\text{ren})}(s,m_{3}^{2},m_{4}^{2}) \im
L^{(\text{ren})}(m_{3}^{2}) \im G^{(\text{vac})}(m_{4}^{2}),
\end{equation}
where due to the renormalization conditions (\ref{16}) the kernel
$K_{2}^{(\text{ren})}$ is given by
\begin{equation}
\label{29}
\begin{split}
  K_{2}^{(\text{ren})}(s,m_{3}^{2},m_{4}^{2}) = &
  K_{1}^{(\text{reg})}(s,m_{3}^{2},m_{4}^{2}) -
  K_{1}^{(\text{reg})}(m^2,m_{3}^{2},m_{4}^{2}) - \\ & - (s-m^2)
  [\partial_s K_{1}^{(\text{reg})}(s,m_{3}^{2},m_{4}^{2})]_{s=m^2}.
\end{split}
\end{equation} 
The cancellation of the contributions from the subtraction of the
sub-divergences is due to a specialty of the sunset-diagram: Here all
contracted diagrams are of tadpole structure and therefore independent of
the external momentum $p$ and thus are completely canceled by the overall
subtractions. The advantage to take them nonetheless into account is that
at any stage of the calculation we use renormalized functions which can be
calculated numerically without using any intermediate regularized
functions.

\begin{figure}
\begin{minipage}{0.49\textwidth}
\centering{\includegraphics[width=\textwidth]{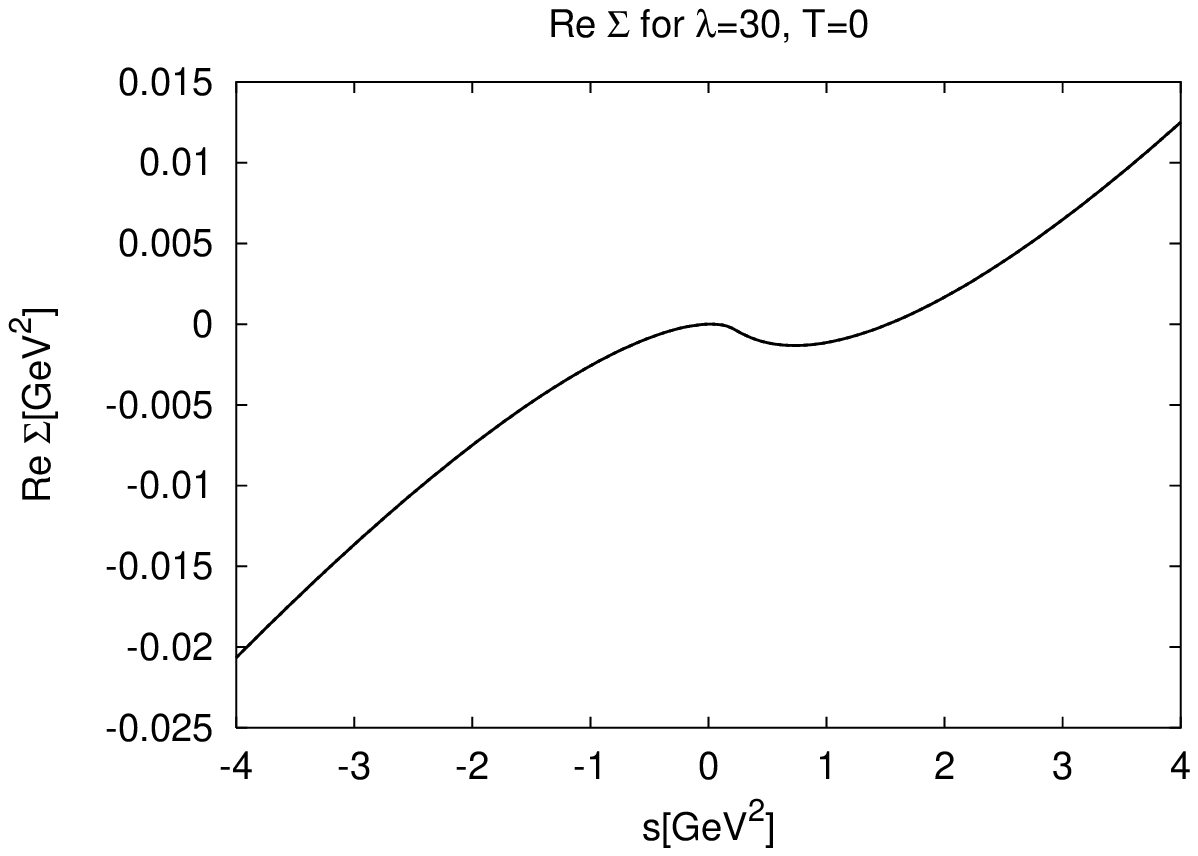}}
\end{minipage}
\begin{minipage}{0.49\textwidth}
\hspace{2mm}
\centering{\includegraphics[width=\textwidth]{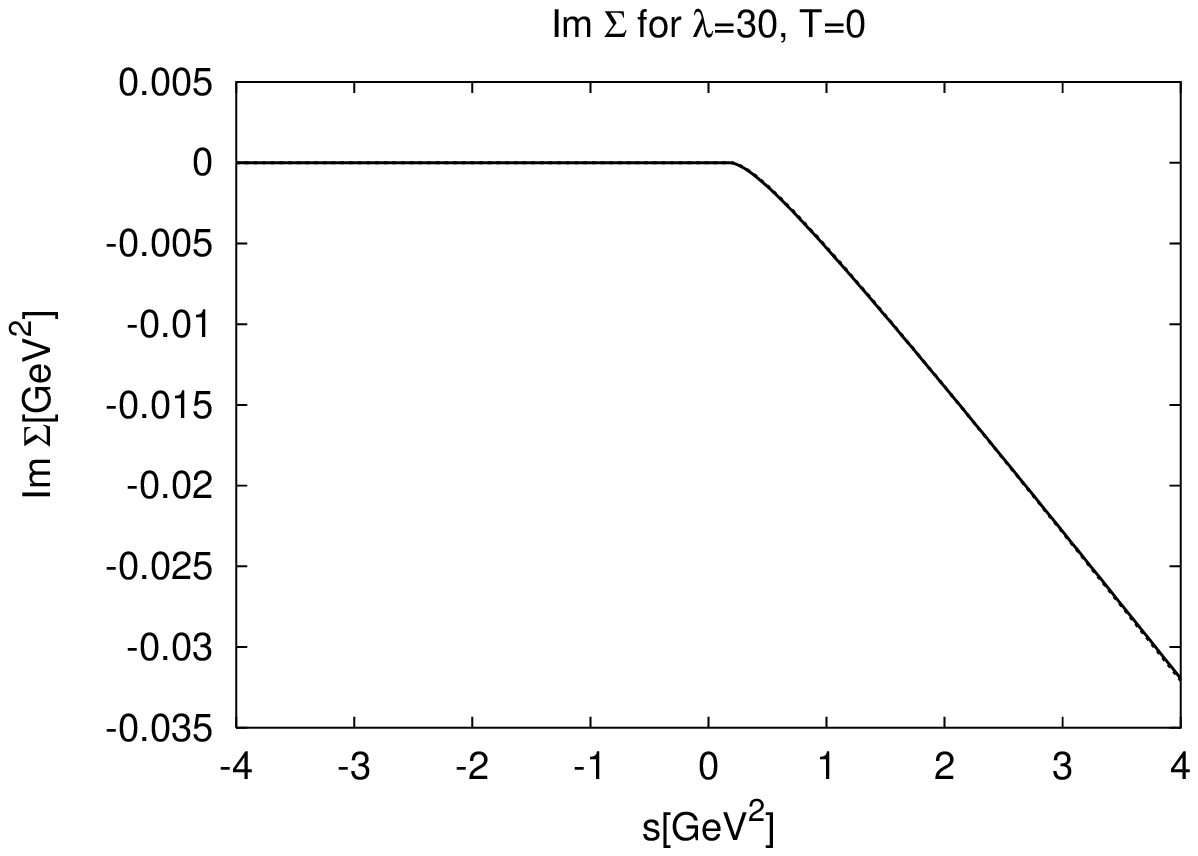}}
\end{minipage}
\caption{Real (left) and imaginary part (right) of the sunset self-
  energy. The perturbative and the self-consistent result lie on top
  of each other due to the large threshold at $s=9 m^2$.}
\label{fig2}
\end{figure}

From (\ref{23}) we find the analytical expression for the kernel $K_{2}$:
\begin{equation}
\label{30}
\begin{split}
  K_{2}^{(\text{ren})}(s,m_{3}^{2},m_{4}^{2})= & \frac{1}{16 \pi^{2}
    m^{2} s} \Bigg \{m^{2} \lambda(s,m_{3},m_{4}) \times \Bigg [
  \artanh \left ( \frac{s+m_{3}^{2}-m_{4}^{2}}{\lambda(s,m_{3},m_{4})}
  \right ) \\ & + \artanh \left (
    \frac{s-m_{3}^{2}+m_{4}^{2}}{\lambda(s,m_{3},m_{4})} \right )
  \Bigg ] \\ & -s \lambda(m^{2},m_{3},m_{4}) \times \Bigg [ \artanh
  \left ( \frac{m^{2}+m_{3}^{2}-m_{4}^{2}}{\lambda(m^{2},m_{3},m_{4})}
  \right ) \\ & + \artanh \left (
    \frac{m^{2}-m_{3}^{2}+m_{4}^{2}}{\lambda(m^{2},m_{3},m_{4})}
  \right )
  \Bigg ] \Bigg \} \\
  & + \frac{
    (s-m^{2})[(m_{3}^{2}-m_{4}^{2})^{2}-m^{2}(m_{3}^{2}+m_{4}^{2})]}{16
    \pi^{2} m^{4} \lambda(m^{2},m_{3},m_{4})} 
  \\
  & \times \Bigg [ \artanh \left (
    \frac{m^{2}+m_{3}^{2}-m_{4}^{2}}{\lambda(m^{2},m_{3},m_{4})}
  \right ) + \artanh \left (
    \frac{m^{2}-m_{3}^{2}+m_{4}^{2}}{\lambda(m^{2},m_{3},m_{4})}
  \right ) \Bigg ] \\ & - \frac{s-m^{2}}{16 \pi^{2}m^{2}} +
  \frac{(m_{3}^{2}-m_{4}^{2}) (s-m^{2})^2}{16 \pi^{2} m^{4} s} \ln
  \left ( \frac{m_{3}}{m_{4}} \right).
\end{split}
\end{equation}
The equations (\ref{20}) and (\ref{28}), supplemented by the Dyson equation
\begin{equation}
\label{31}
G^{(\text{vac})}(p)=\frac{1}{p^{2}-m^{2}-\Sigma^{(\text{vac})}(p^{2})+\ii
    \eta},
\end{equation}
form the closed set of renormalized equations of motion, which were solved
with the analytically given Kernels $K_{1}^{(\text{ren})}$ and
$K_{2}^{(\text{ren})}$ from (\ref{25}) and (\ref{30}). For the integrals
(\ref{20}) and (\ref{28}) a simple adaptive Simpson integrator was used. We
have chosen $m=m_{\pi}=140 \, \text{MeV}$ for the mass and $\lambda=30$. As
Fig. \ref{fig2} shows for this coupling there is no visible difference
between the perturbative and the self-consistent result since the main
contribution comes from the pole term of the propagator, while the
continuous part, which starts at a threshold of $s=9 m^2$, is suppressed.
Thus, due to our renormalization scheme, where $m$ is the physical mass
parameter, the pole-term result essentially coincides with the perturbative
one.

\subsection{The Bethe-Salpeter function}

The renormalized Bethe-Salpeter kernel $\Gamma^{(4, \text{ren})}$
defined in (\ref{4}) reads 
\begin{equation}
\label{32}
\Gamma^{(4,\text{ren})}(l,p) := \Gamma^{(4,\text{ren})}[(l-p)^2] = 
\frac{\lambda}{2}+\lambda^2 L^{(\text{ren})}[(l-p)^2]. 
\end{equation}
Since as shown above the pole term essentially determines the
loop-function $L$, we simplified the task to solve (\ref{5}) by using
the free propagator 
\begin{equation}
\Delta(p^{2})=\frac{1}{p^{2}-m^{2}+\ii \eta},
\end{equation}
instead of the self-consistent vacuum $G^{(\text{vac})}$.

Subtracted dispersion relations for
$\Gamma^{(4,\text{ren})}$ and $\Lambda^{(\text{ren})}$ and the
renormalization conditions
$\Gamma^{(4,\text{ren})}(0,0)=\Lambda^{(\text{ren})}(0,0)=\lambda/2$
provide the renormalized integral equation
\begin{equation}
\label{33}
\begin{split}
  \Lambda^{(\text{ren})}(0,p^2)= & \Gamma^{(4,\text{ren})}(p^2)+ \ii
  \fint{l} \left [\Delta(l^2) \right]^2 \times \\
 & \times \Bigg \{ \int_{4m^2}^{\infty} \frac{\d (m_{2}^{2})}{\pi} \im
  \Gamma^{(4,\text{ren})}(m_{2}^2) \frac{2 l p -
    p^2}{[m_{2}^2-(l-p)^2-\ii \eta](m_{2}^2-l^{2}-\ii \eta)} \times
  \\ & \times \left [\int \frac{\d (m_{1}^{2})}{\pi m_{1}^{2}} \frac{l^2
      \im \Lambda^{(\text{ren})}(0,m_{1}^2)}{m_{1}^{2}-l^{2} - \ii
      \eta} + \frac{\lambda}{2} \right] \Bigg \}.
\end{split}
\end{equation}
Again the $l$-integration can be performed with the help of
standard perturbation theory integrals leading to the kernels
\begin{equation}
\label{34}
\begin{split}
  & K_{3}^{(\text{ren})}(p^2,m_{1},m_{2}) = \ii \fint{l}
  \frac{1}{(m^2-l^2-\ii \eta)^2} \frac{l^2(2 l p
    -p^2)}{(m_{1}^{2}-l^{2}-\ii \eta)[m_{2}^{2} -
    (l-p)^2-\ii \eta]}, \\
  & K_{4}^{(\text{ren})}(p^2,m_{2})=\ii \fint{l} \frac{2 l
    p-p^2}{(m^{2}-l^{2} -\ii \eta)^2[m_{2}^{2}-(l-p)^{2}-\ii
    \eta]},
\end{split}
\end{equation}
which can be related to the previous kernel $K_1$ through the identity
\begin{equation}
\label{35}
[\Delta(l^{2})]^{2} = -\frac{1}{2m} \partial_{m}
\frac{1}{m^{2}-l^{2}-\ii \eta}.
\end{equation}
We start with the calculation of $K_{4}$ defined in (\ref{34}) and
write it in the form
\begin{equation}
\label{36}
\begin{split}
  K_{4}(s,m,m_{2}) = -\frac{\ii}{2m} \partial_{m} \fint{l}
  \frac{1}{m^2-l^2-\ii \eta} \Bigg [ &
  \frac{1}{m_{2}^{2}-(l-p)^{2}-\ii \eta} -
  \frac{1}{m_{2}^{2}-l^{2}-\ii \eta} \Bigg ]
\end{split}
\end{equation}
which can be expressed in terms of $K_{1}^{(\text{ren})}$
\begin{equation}
\label{37}
\begin{split}
  K_{4}(s,m,m_{2}) & = -\frac{1}{2m} \partial_{m}
  K_{1}(s,m,m_{2}) = -\frac{1}{16 \pi^2 s} \Bigg \{
  \frac{s}{m_{2}^{2}-m^{2}}
  + \frac{(m^{2}-m_{2}^{2}-s)}{\lambda(s,m,m_{2})} \\
  & \quad \times \left [ \artanh \left
      (\frac{s+m_{1}^2-m_{2}^2}{\lambda(s,m_{1},m_{2})} \right ) +
    \artanh \left (\frac{s-m_{1}^2+m_{2}^2}{\lambda(s,m_{1},m_{2})}
    \right ) \right
  ] \\
  & \quad + \frac{(m^{2}-m_{2}^{2})^{2}+2 m_{2}^{2}
    s}{(m^{2}-m_{2}^2)^2} \ln \left (\frac{m}{m_{2}} \right) \Bigg
  \}.
\end{split}
\end{equation}
The integrand of $K_{3}$ cf. (\ref{34}) can
be rewritten as follows
\begin{equation}
\label{38}
\begin{split}
  \frac{\ii}{m^{2}-l^{2}-\ii \eta} \frac{l^{2}}{m_{1}^{2}(m_{1}^{2} -
    l^{2} -\ii \eta)} \frac{p^{2} - 2 l p}{(m_{2}^{2}-l^{2}-\ii
    \eta)[m_{2}^{2} - (l-p)^{2}-\ii \eta]} = \\
  = -\frac{\ii}{2m} \partial_{m} \frac{1}{m_{1}^{2}-m^{2}} \left (
    \frac{1}{m^{2}-l^{2}-\ii \eta} - \frac{1}{m_{1}^2-l^{2}-\ii
      \eta} \right ) - \\
  - \frac{1}{m_{1}^{2}(m^{2}-l^{2}-\ii \eta)} \left [
      \frac{1}{m_{2}^{2} - (l-p)^{2} - \ii \eta} - \frac{1}{m_{2}^{2}
        - l^{2} - \ii \eta} \right ].
\end{split}
\end{equation}
The integration of (\ref{38}) then leads to
\begin{equation}
\label{39}
K_{3}(s,m,m_{1},m_{2})=\frac{K_{1}(s,m_{1},m_{2}) -
  K_{1}(s,m,m_{2})}{(m^{2}-m_{1}^{2})^{2}} + 
\frac{m^{2}-2m_{1}^{2}}{(m^{2}-m_{1}^{2})m_{1}^{2}} K_{4}(s,m,m_{2}).
\end{equation}
With the so defined kernels we can express (\ref{33}) as follows:
\begin{equation}
\begin{split}
  \Lambda^{(\text{ren})}(0,p^2)=\Gamma(p^2) & + \int_{4 m^2}^{\infty}
  \frac{\d m_{1}^2}{\pi m_{1}^2} \frac{\d m_{2}^{2}}{\pi}
  K_{3}(p^2,m_{1},m_{2}) \im \Lambda^{(\text{ren})}(0,m_{1}^2) \im
  \Gamma^{(4,\text{ren})}(m_{2}^{2}) +\\ &+ \frac{\lambda}{2} \int_{4
    m^2}^{\infty} \frac{\d m_{2}^{2}}{\pi} K_{4}(p^2,m_{2}) \im
  \Gamma^{(4,\text{ren})}(m_{2}^2).
\end{split}
\end{equation}
\begin{figure}
\begin{minipage}{0.49\textwidth}
\centering{\includegraphics[width=\textwidth]{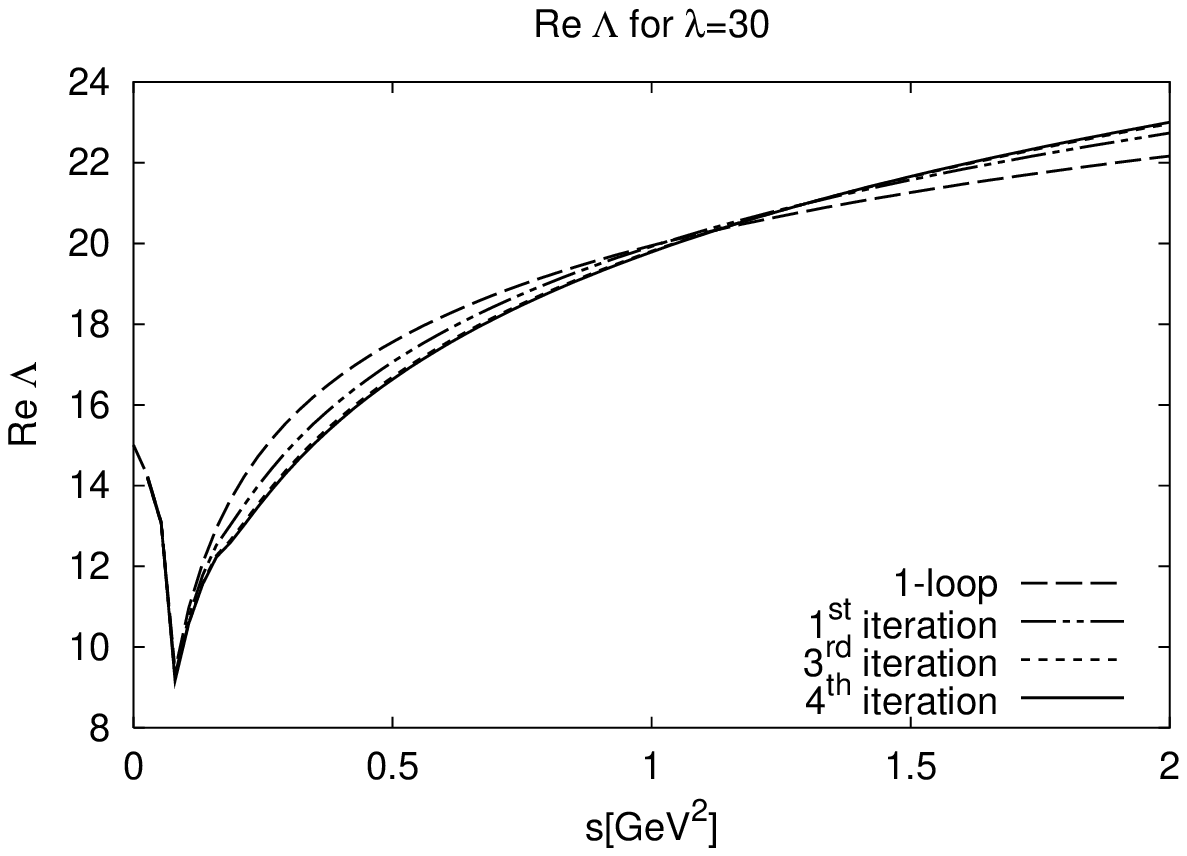}}
\end{minipage}\hspace{2mm}
\begin{minipage}{0.49\textwidth}
\centering{\includegraphics[width=\textwidth]{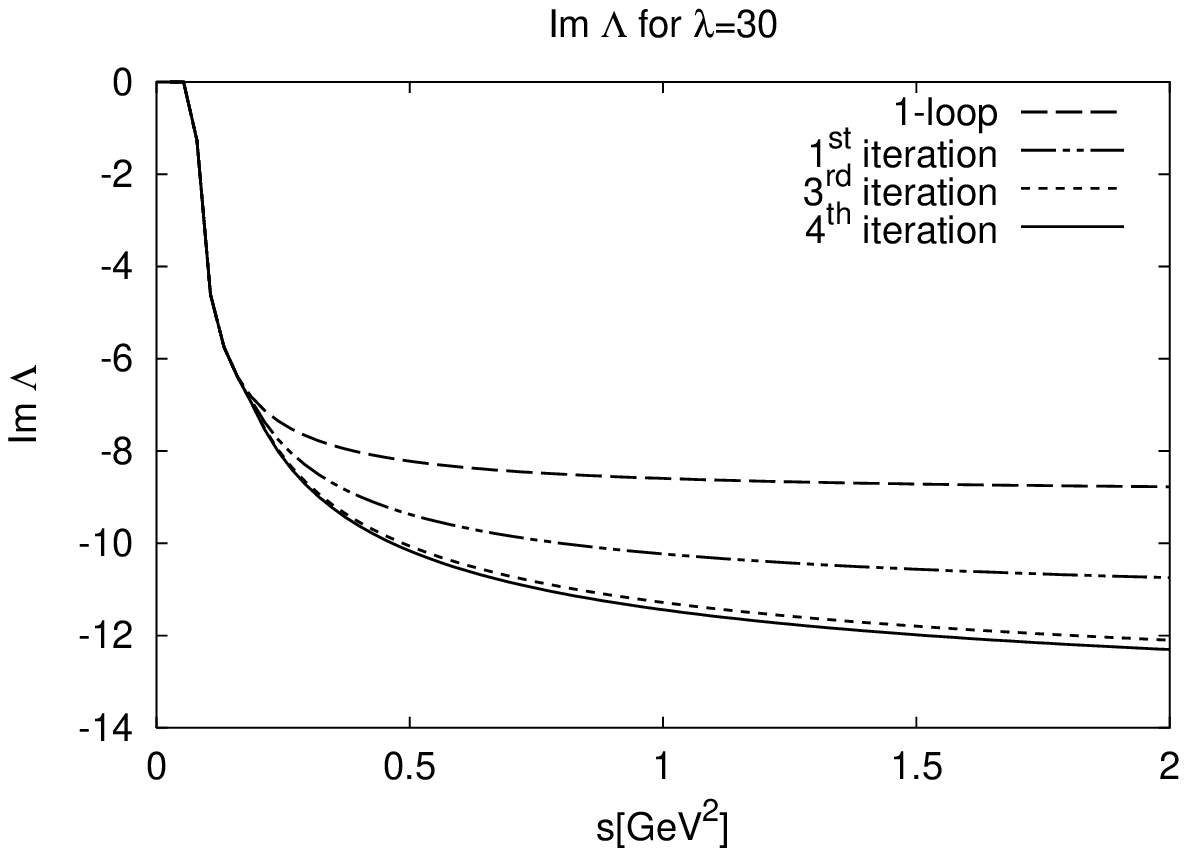}}
\end{minipage}
\caption{Real (left) and imaginary part (right) of the
  $\Lambda$-function. The plot shows that after three iterations the
  solution is already stable.}
\label{fig3}
\end{figure}
Fig. \ref{fig3} shows the solution of this renormalized equation of motion
in comparison with the one-loop approximation $\Gamma^{(4,\text{ren})}$. As
the figure shows, the solution is stable after three iterations.

\subsection{The temperature dependent parts}

To calculate the temperature dependent part of the self-energy we use
the analytical properties of the two-point functions and the fact that
the retarded propagator fulfills the simple Dyson equation
\begin{equation}\label{G_R}
G_R(p)=\frac{1}{p^2-m^2-\Sigma_R(p)},
\end{equation}
where the retarded self-energy can be obtained from the
$\{--\}$-matrix element via the equilibrium expression
\begin{equation}
\Sigma_{R}(p)=\re \Sigma^{--}(p)+\ii \tanh \left ( \frac{p_{0}}{2T}
\right ) \im \Sigma^{--}(p).
\end{equation}
As explained in I both, $G^{(\text{vac})}$ and $\Sigma^{(0)}$, are
diagonal matrices while the corresponding off-diagonal parts are
contained in $G^{(\text{r})}$, because they are of a lower degree of
divergence due to explicit $\Theta+n$-factors.

Due to its topology the sunset-diagram has no further vertices
besides its two external points. Therefore the 2PI vacuum four-point function
$\Gamma^{(4,\text{vac})}$ is effectively only a two-point function and
thus we need only the $\{--\}$-components of the various Green's
functions in order to calculate $\Sigma^{--}$. This  implies
\begin{equation}
G^{--(\text{matter})}=G^{--}-G^{--(\text{vac})}, \quad
G^{--(\text{r})}=G^{--(\text{matter})}-[G^{--(\text{vac})}]^2
\Sigma^{--(0)}.
\end{equation}
With help of the renormalized vacuum parts we obtain for the
self-energy
\begin{equation}\label{Sigma--}
\begin{split}
  \Sigma^{--}(p)= & \Sigma^{--(\text{vac})}(p)+ \\
  & + \ii \fint{l} \left \{
    \Gamma^{(4,\text{vac})}[(l+p)^2]-\Gamma^{(4,\text{vac})}(l^2)
  \right \} G^{--(\text{matter})}(l) + \\
  & + \ii \fint{l} \Lambda^{(\text{ren})}(0,l^2)
  G^{--(\text{r})}(l) - \\
  & - \frac{\lambda^{2}}{2} \fint{l_{1}} \fint{l_{2}}
  G^{--(\text{matter})}(l_1) G^{--(\text{matter})}(l_1+l_2)
  G^{--(\text{vac})}(l_2+p) - \\
 & -\frac{\lambda^{2}}{6} \fint{l_{1}} \fint{l_{2}}
  G^{--(\text{matter})}(l_1) G^{--(\text{matter})}(l_1+l_2)
  G^{--(\text{matter})}(l_2+p).
\end{split}
\end{equation}
Note that all vacuum quantities entering here are to be taken in their
\emph{renormalized} version and that the remaining integrals are all
finite due to power counting.

\begin{figure}
\centering{
\begin{minipage}{0.49\textwidth}
\centering{\includegraphics[width=\textwidth]{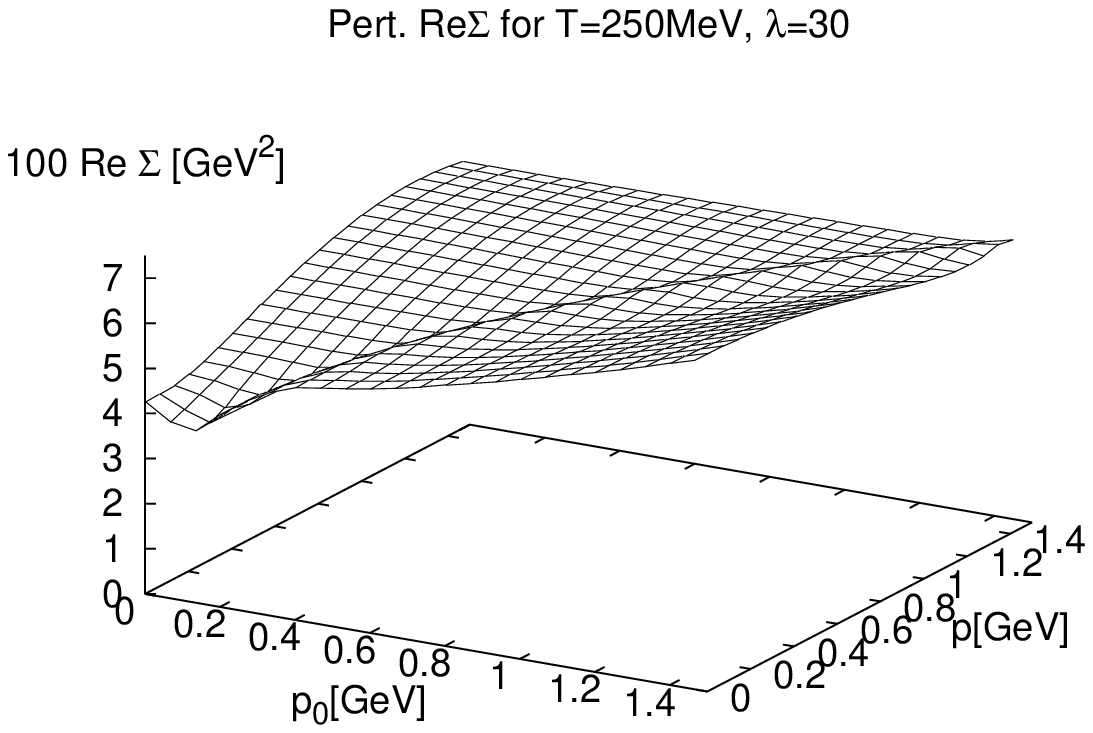}}
\end{minipage}\hspace*{0.2cm}
\begin{minipage}{0.49\textwidth}
\centering{\includegraphics[width=\textwidth]{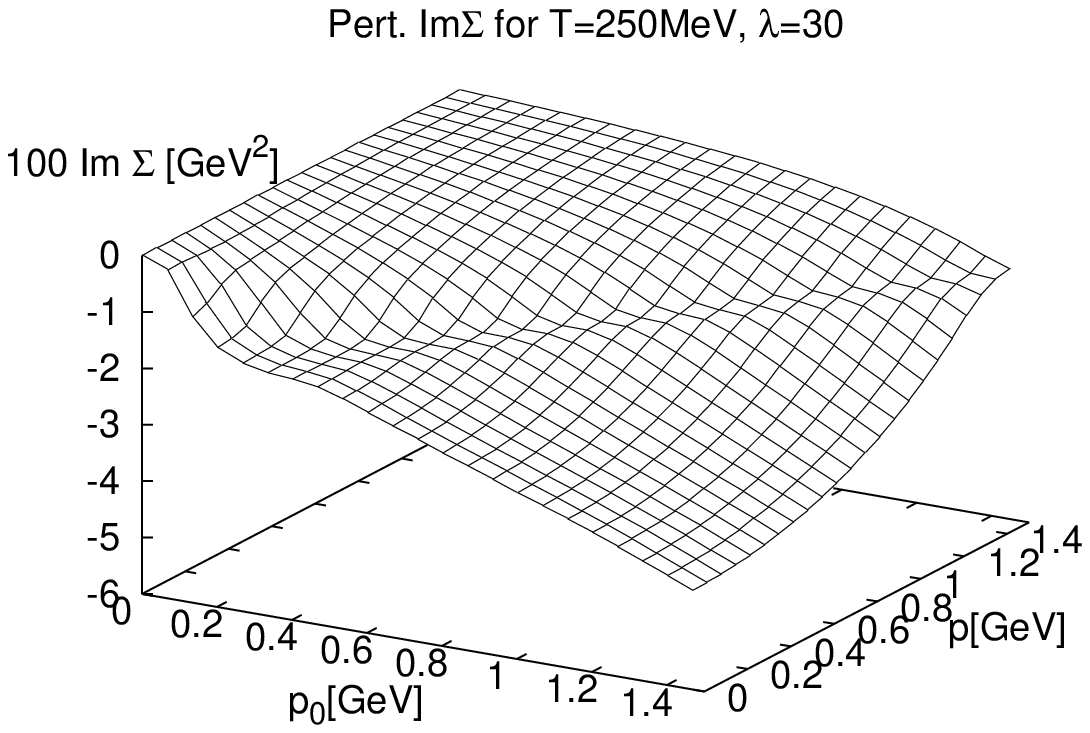}}
\end{minipage}}
\vspace{10mm}
\centering{
\begin{minipage}{0.49\textwidth}
\centering{\includegraphics[width=\textwidth]{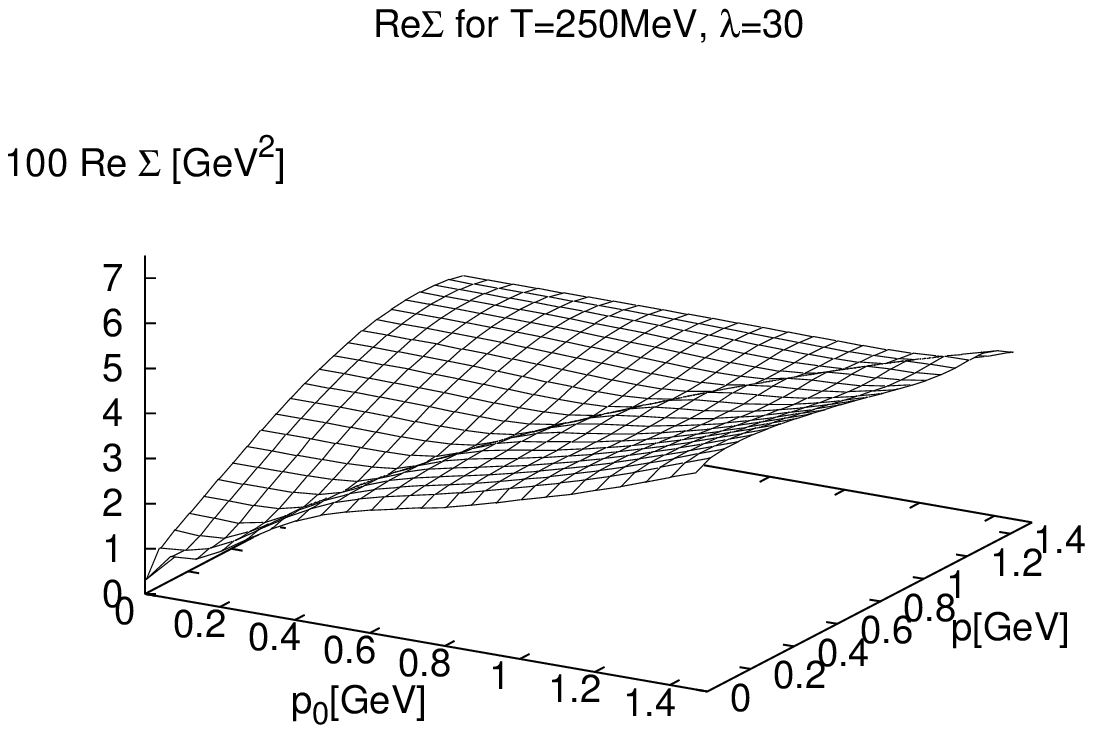}}
\end{minipage}\hspace*{0.5cm}
\begin{minipage}{0.49\textwidth}
\centering{\includegraphics[width=\textwidth]{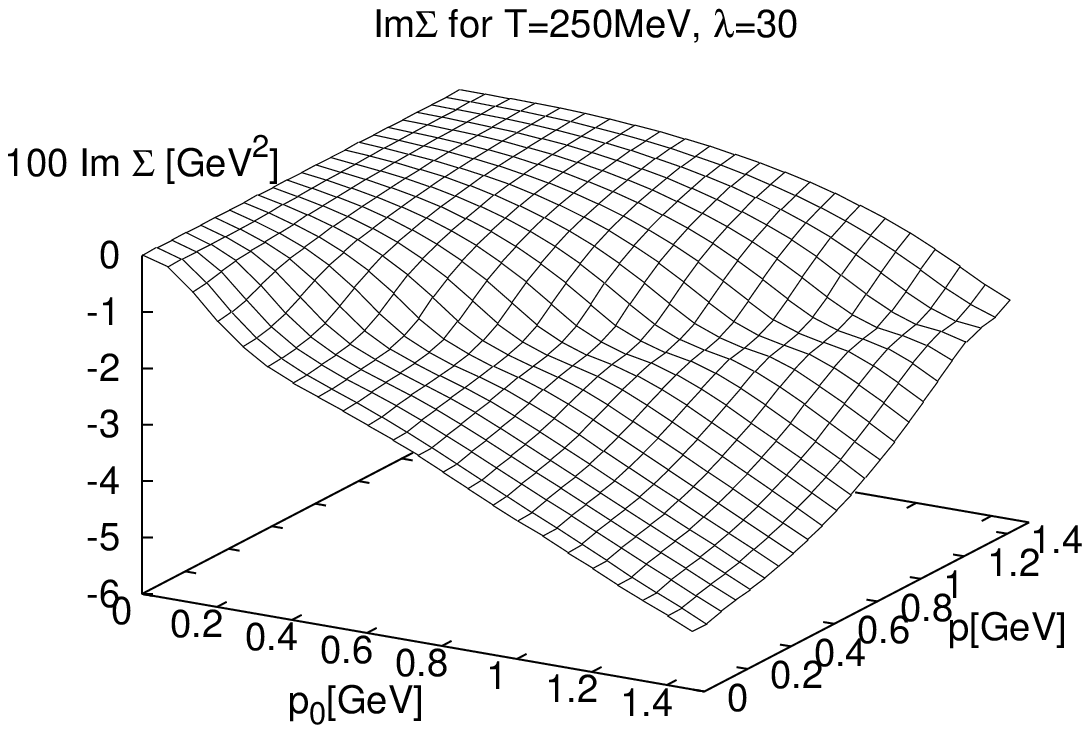}}
\end{minipage}
}
\caption{Real (left) and imaginary part (right) of the perturbative
  (top) and the self-consistent self-energy for $\lambda=30$,
  $m=140\text{MeV}$ and $T=250 \text{MeV}$. Note that the self-energies are
  multiplied with an factor $100$ in these plots!}
\label{fig4} 
\end{figure}
\subsection{Results}
The above finite integrals are to be evaluated numerically. While due to
Lorentz invariance the vacuum loops involve just two-dimensional
integrations which can numerically be integrated by standard methods, for
the in-matter loops only rotational symmetry in three momentum space can be
exploited. For each loop diagram this leads to three dimensional integrals
for each external momentum $p_0,|\vec{p}|$.  We solve these integrals on an
equal distant $N\times N$ lattice in these coordinates.  Naively the
computing effort would then scale like $N^5$. However, we succeeded to
develop an improved algorithm for the loop integrations, where the
computing time scales with $N^4$ (essentially a gain of more than two
orders of magnitude).  The lattice implies an infrared cut-off which
requires particular care for the treatment of the sharp structures of the
vacuum propagator around its on-shell pole. The method at hand was to
tabulate the lattice cell integrated values of the propagator and its
moment in $p_0$ given by the analytic result from a linear interpolation in
$p_0^2$ of the inverse propagator at fixed $|\vec{p}|$. The remaining
self-energy factors of the integrands are approximated to linear order in
$p_0^2$.  The interpolation procedure is adapted to the fact that due to the
quadratic pole of the $(G^{--\text{(vac)}})^2$ term one is sensitive to the
$p_0$ derivative (residue) of the remaining integrand. Both, the infrared
and the UV cut-offs of the lattice have been controlled by varying the
lattice spacing and the overall size of the lattice. The final results were
achieved with $\Delta p_0=\Delta |\vec{p}|=6.67 \,\text{MeV}$ with $N=300$,
i.e., a UV-cutoff at $2\,\text{GeV}$.

\begin{figure}
\centering{
\begin{minipage}{0.49\textwidth}
\centering{\includegraphics[width=\textwidth]{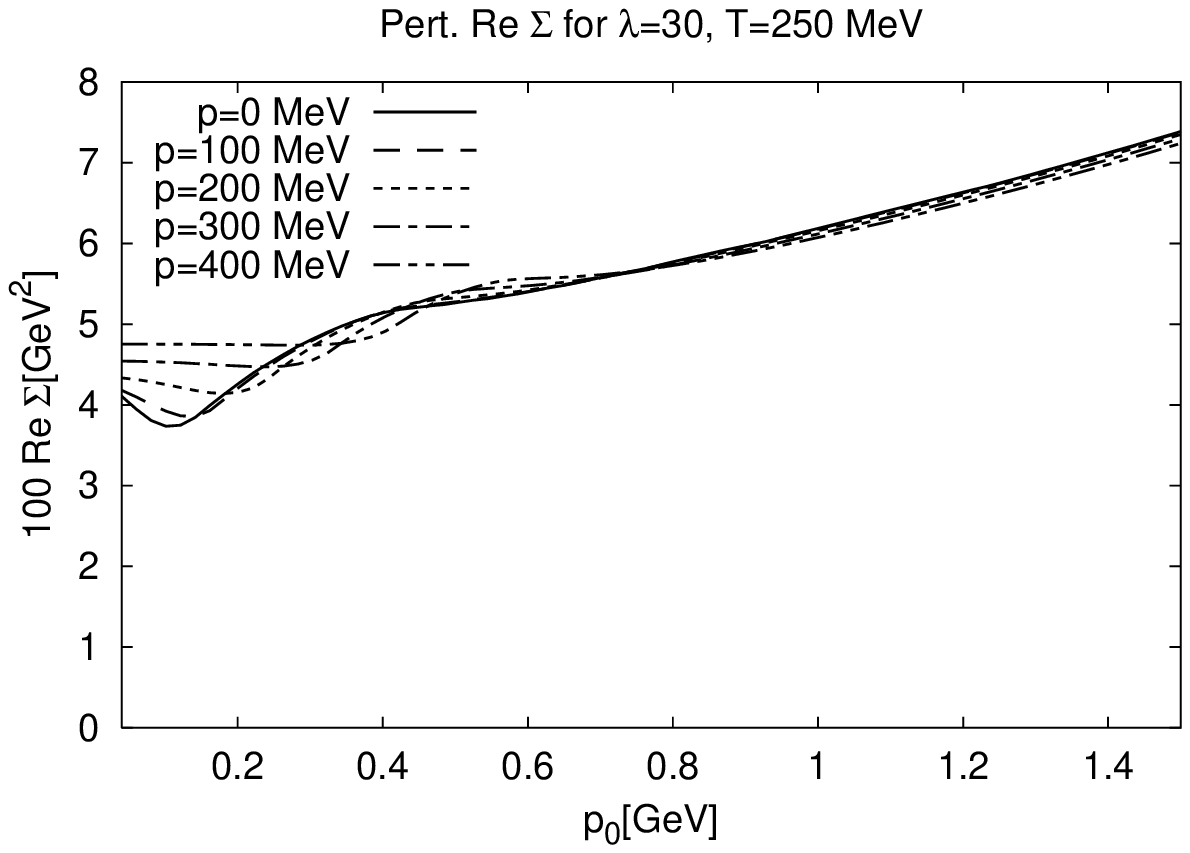}}
\end{minipage}\hspace*{0.2cm}
\begin{minipage}{0.49\textwidth}
\centering{\includegraphics[width=\textwidth]{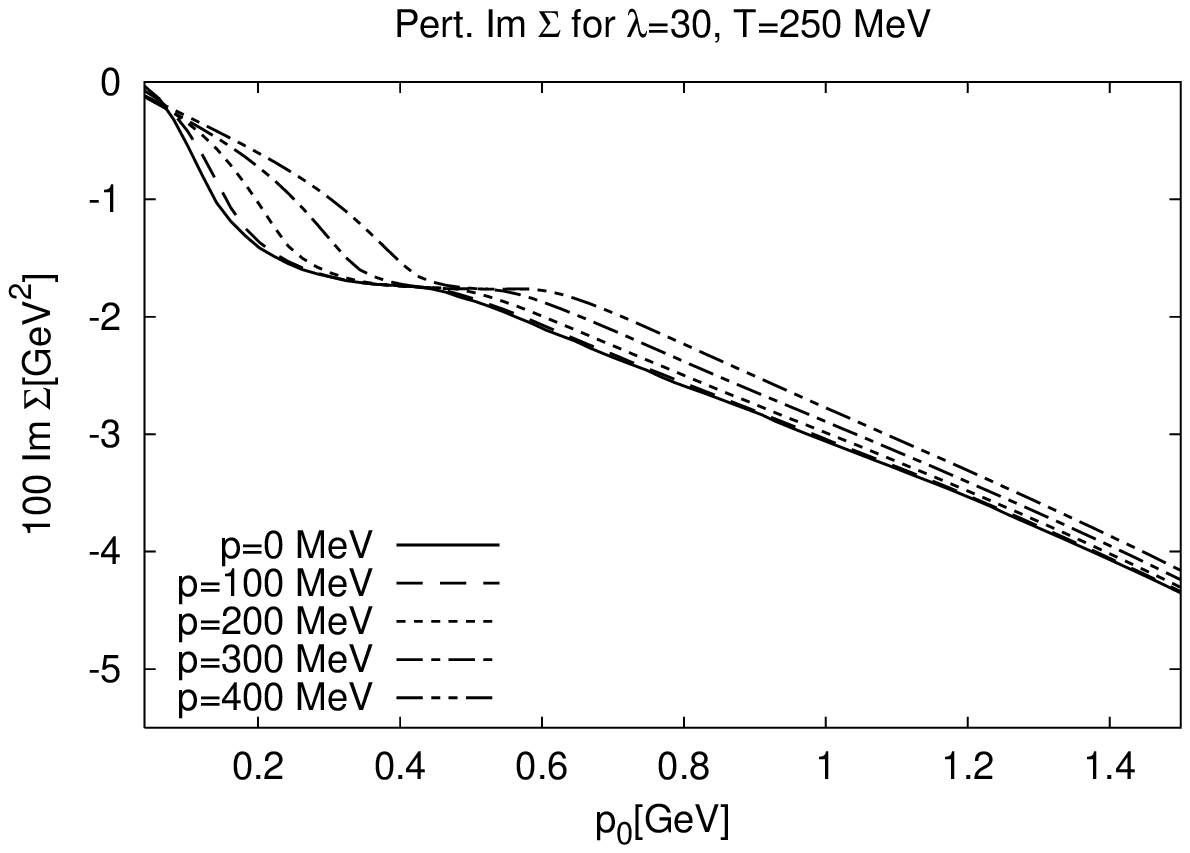}}
\end{minipage}}
\vspace{10mm}
\centering{
\begin{minipage}{0.49\textwidth}
\centering{\includegraphics[width=\textwidth]{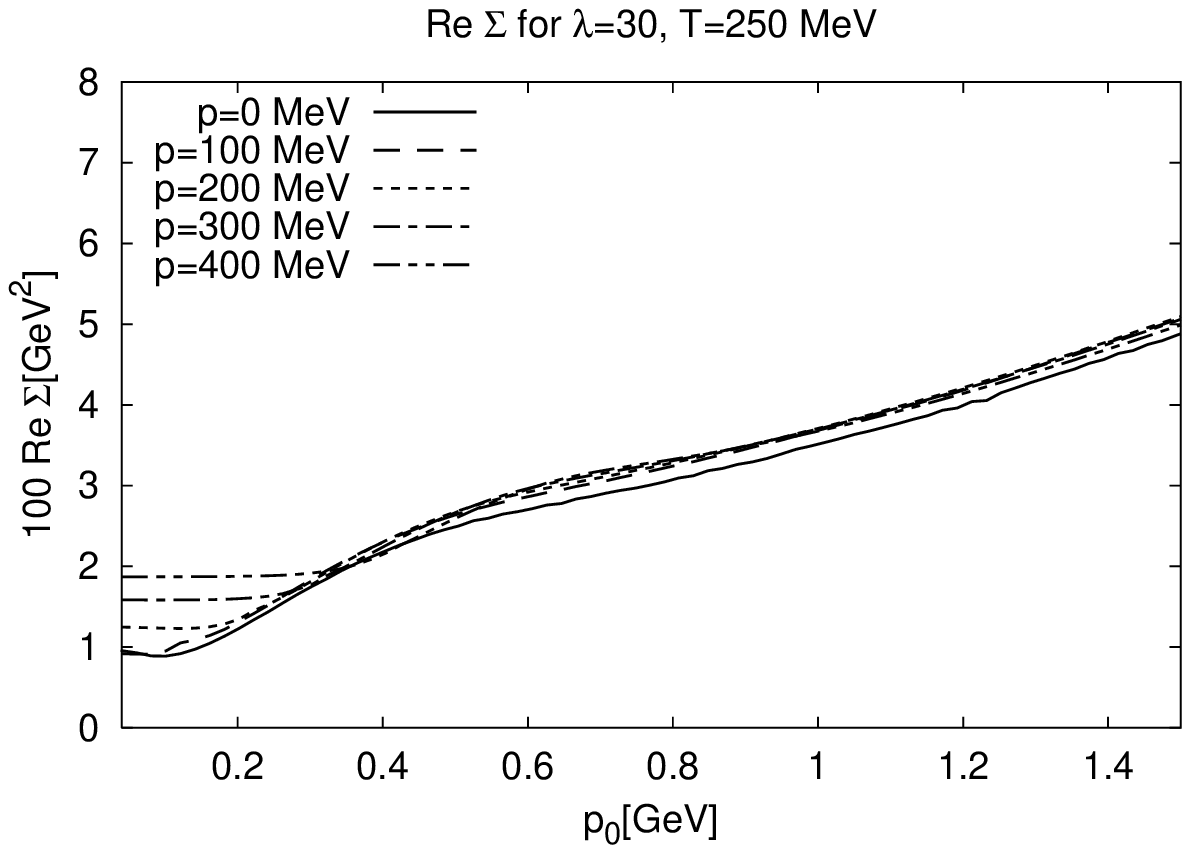}}
\end{minipage}\hspace*{0.2cm}
\begin{minipage}{0.49\textwidth}
  \centering{\includegraphics[width=\textwidth]{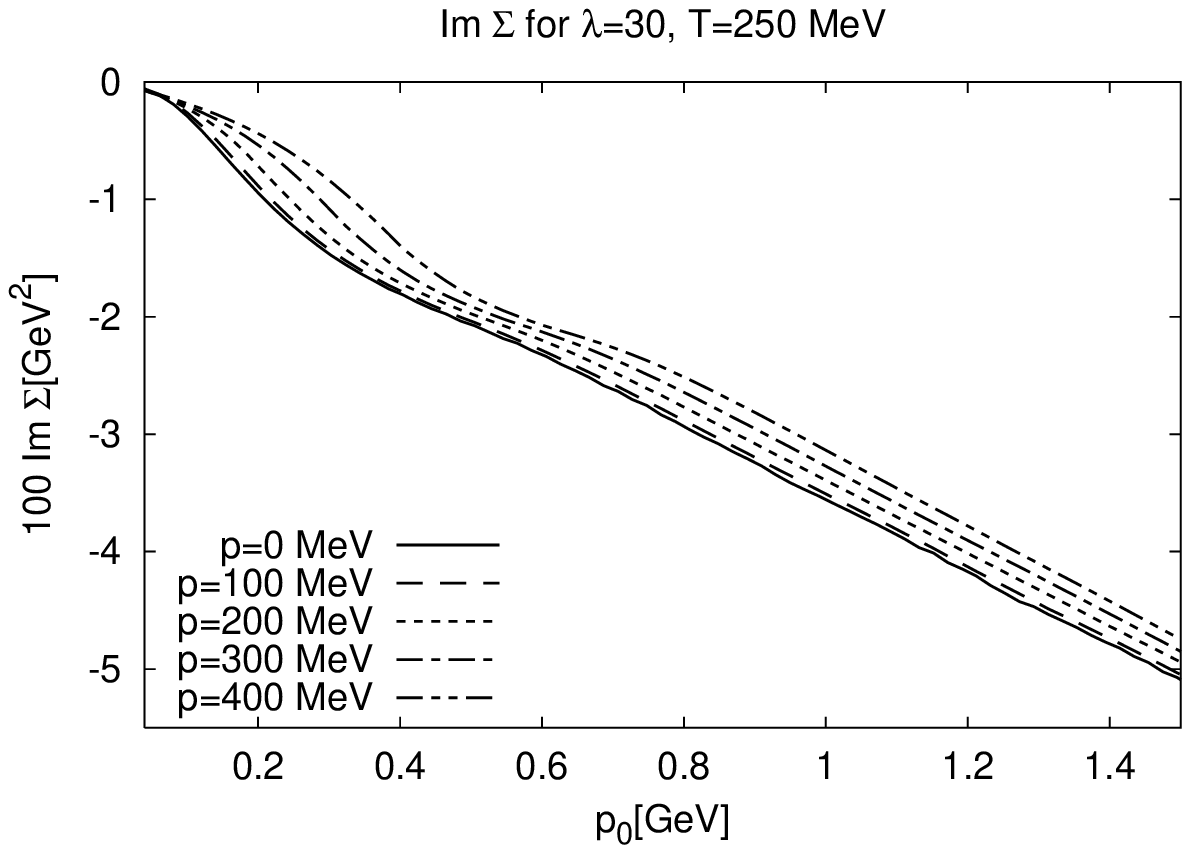}}
\end{minipage}
}
\caption{Real (left) and imaginary part (right) of the perturbative
  (top) and the self-consistent self-energy for $\lambda=30$,
  $m=140\text{MeV}$ and $T=250\,\text{MeV}$ as a function of $p_0$ for
  various 3-momenta. Note that the self-energies are multiplied by a factor
  of $100$ in these plots!}
\label{fig5} 
\end{figure}

Perturbative results are obtained through Eq. (\ref{Sigma--}) approximating
$G^{\text{(matter)}}$ by the KMS-temperature part of the free propagator
$G^{(\text{vac};T)}$ (last term in Eq.  (\ref{GvacT}); note that here
$G^{\text{(r)}}=G^{\text{(matter)}}$ since $\Sigma^{\text{(vac)}}=0$). The
self-consistent solutions are then obtained iteratively through the set of
Eqs. (\ref{G_R}) to (\ref{Sigma--}). The results for both, the perturbative
and the self-consistent case, are shown in Fig. \ref{fig4} in a
3-dimensional plot over the $(p_0,|\vec{p}|)$-plane illustrating that the
entire calculations are performed with the full dependence on energy and
momentum. Details can be extracted from the cuts shown for a set of
selected momenta in Fig. \ref{fig5}.
The main qualitative results are similar for both the perturbative and the
self-consistent calculation: In the vacuum and self-consistent pure tadpole
case the self-energy shows a threshold cut resulting from the decay into
three particles, i.e., $p_0^2-\vec{p}^2\ge 9M^2$.  Adding the sunset
self-energy leads to a spectral width which dissolved this threshold such
that the self-energy shows spectral strength (imaginary parts) at all
energies.  While the growing high-energy tail is related to the decay of
virtual bosons into three particles, at finite temperature, as a new
component, a low-energy plateau in $\im \Sigma^{\text{R}}$ emerges from
in-medium scattering processes. 

Various balancing effects are encountered for the self-consistent case: For
sufficiently large couplings and/or temperatures the self-consistent
treatment shows quantitative effects on the width. The finite spectral
width itself leads to a further broadening of the width and a smoothing of
the structures as a function of energy.
\begin{figure}
\centering{
\begin{minipage}{0.49\textwidth}
\centering{\includegraphics[width=\textwidth]{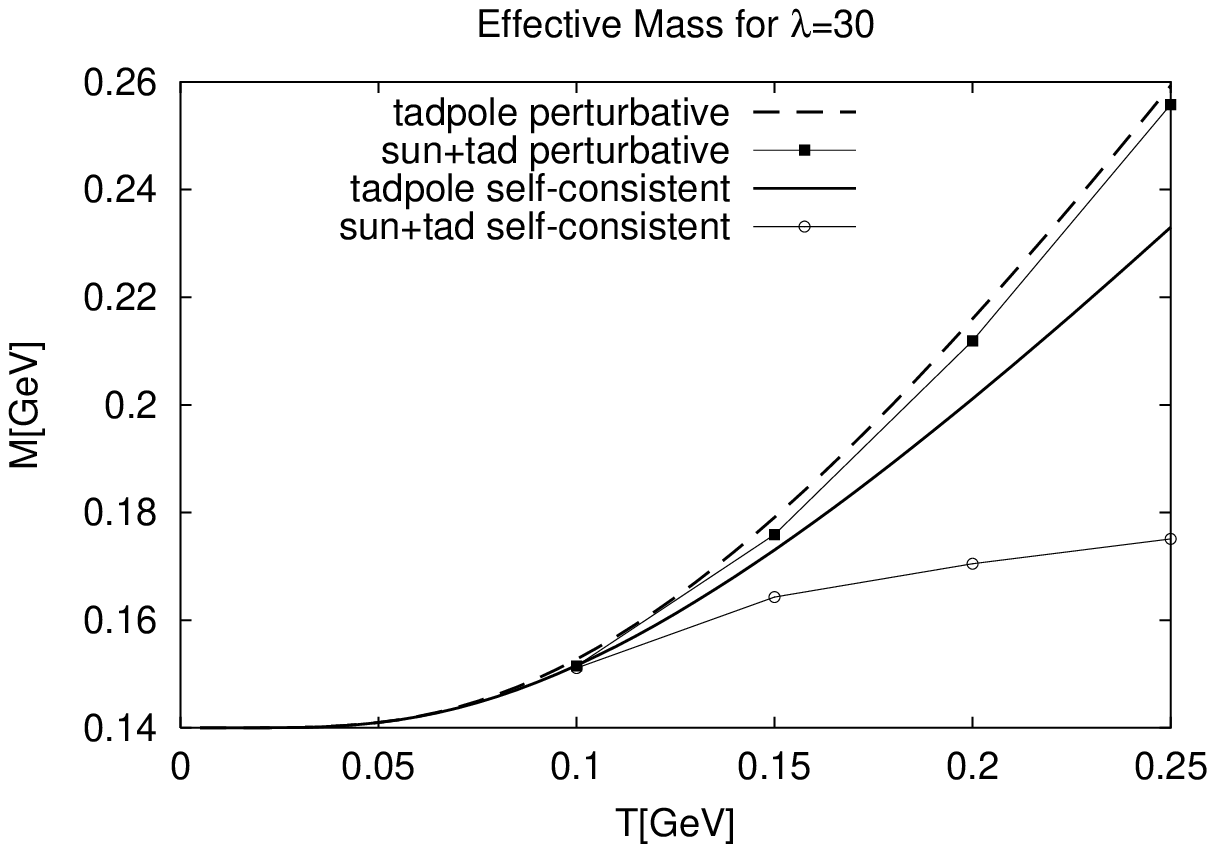}}
\end{minipage}\hspace*{0.2cm}
\begin{minipage}{0.49\textwidth}
\centering{\includegraphics[width=\textwidth]{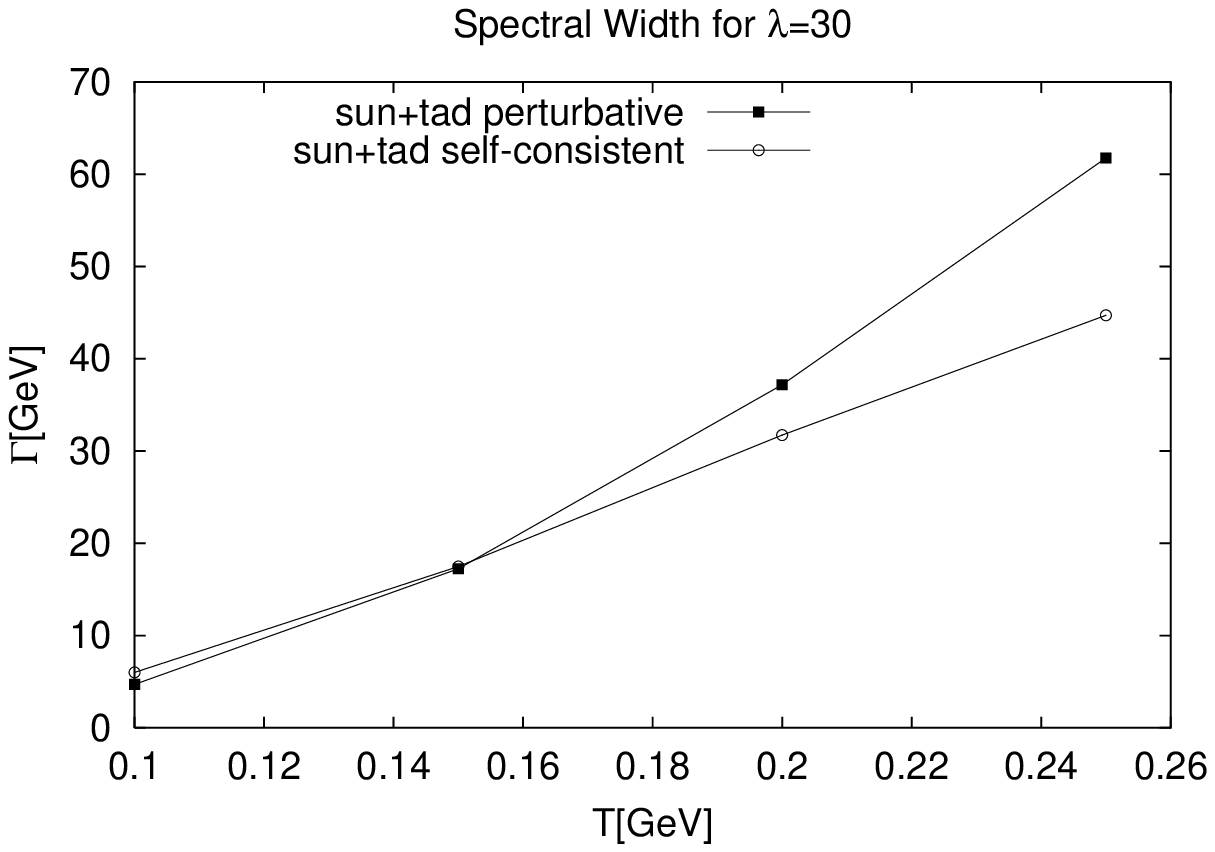}}
\end{minipage}}
\caption{The in-medium effective masses $M$ (left) and spectral widths
  $\Gamma$ (right) of the particles for the various approximations
  described in the text as a function of the system's temperature $T$.}
\label{fig6}
\end{figure}
This is however counter balanced by the behavior of the real part of the
self-energy, which, as discussed below, essentially shifts the in-medium
mass upwards. This reduces the available phase space for real processes.
With increasing coupling strength $\lambda$ a nearly linear behavior of
$\im\Sigma^{\text{R}}$ with $p_0$ results implying a nearly constant
damping width given by $-\im\Sigma^{\text{R}}/p_{0}$.

The overall normalization of the real part of $\Sigma^{\text{R}}$ is
determined by the renormalization procedure. In this case there are three
counter balancing effects. First the tadpole loop shifts the mass to higher
values. As the tadpole is less effective for higher masses this effect
weakens itself in the self-consistent tadpole treatment, c.f. Figs.
\ref{fig-1} or \ref{fig6}. However, since the sunset part adds spectral
width, it indirectly contributes to the tadpole loop. Since spectral
strength at the lower mass side carries higher statistical weights, the
tadpole loop in turn leads to a further increase of the mass shift, c.f.
the perturbative calculations of sunset \& tadpole in Fig. \ref{fig6}.

The direct contributions of the sunset terms to the real part of the
self-energy become relevant at higher couplings and temperatures. Then the
self-consistency leads to significant effects which contributes to a net
downshift of the real part of the self energy or in-medium mass $M$. The
latter effect finally overrules the tadpole shift and indeed leads to an
overall negative mass shift compared to the (tadpole dominated)
perturbative result.  These effects are illustrated in Fig.  \ref{fig6}
where the in-medium effective mass $M$ and width $\Gamma$ of the
corresponding ``quasi-particles'' are plotted against the temperature.
Thereby $M$ and $\Gamma$ are defined as the quasi-particle energy $M=p_0$
at vanishing real part of the dispersion relation $[p_0^{2}-m^2 -\re
\Sigma(p_{0},\vec{p})]_{p=(M,\vec{0})}=0$ for $\vec{p}=0$, and through
$\Gamma=-\im \Sigma(p)/p_{0}|_{p=(M,\vec{0})}$, respectively.

\section{Conclusions and outlook}
\label{sect-conc}

We have shown that it is possible to use the renormalization scheme,
proposed in \cite{vHK2001-Ren-I}, for numerical investigations of the
self-consistent approximations for the self-energy derived from the
truncated effective action formalism on the 2PI level. Thereby it is very
important to isolate the divergent vacuum parts consistently, in particular
the implicit or hidden ones, from all convergent and in particular
temperature or more generally matter-dependent parts. This could be
provided by the ansatz given in I for both, the propagator and the
self-energy. The renormalized vacuum pieces are obtained using the Lehmann
representation for all two-point quantities. The resulting integration
kernels can then be renormalized by standard technics. The procedure solely
rests on Weinberg's power counting theorem, i.e., on an analysis of the
asymptotic behavior of the propagators.

In this way both, the renormalized vacuum self-energy and four-point
functions can directly be obtained from finite equations. For the finite
temperature parts it is important that they have to be completely excluded
from the counter term structures. This is achieved by the technique developed
in I. Exploratory calculations are shown for the symmetric Wigner-Weyl
phase of the $\lambda\phi^4$ model for the self-consistent treatment of
both the tadpole and the sunset diagram at finite temperature.

The results promise that the method, which is conserving \cite{baym62,kv97}
and thermodynamically consistent, can also be applied for the genuine
non-equilibrium case, i.e., in quantum transport \cite{ikv99-2} or for the
solution of the renormalized Kadanoff-Baym equations. Numerical studies for
non-equilibrium cases (Gaussian initial conditions, spatially homogeneous
systems) were already performed in \cite{berges00a,berges01b,berges01a}.  These
investigations were undertaken in 1+1 dimensions. Our renormalization
scheme should also be applicable for the non-equilibrium case in 1+3
dimensions and implementable in numerical codes.

The investigation of the symmetry properties of $\Phi$-derivable
approximations is the subject of a forthcoming publication
\cite{vHK2001-Ren-III}. It is known that in general the symmetries of the
classical action which lead to Ward-Takahashi identities for the proper
vertex-functions are violated for the self-consistent Dyson resummation for
the functions beyond the one-point level, i.e., on the correlator level.
The reason is that, although the \emph{functional} $\Gamma$ can be expanded
with respect to expansion parameters like the coupling or $\hbar$ (loop
expansion) or large-$N$ expansions for O($N$) type models, the solution of
the self-consistent equations of motion contains partial contributions to
any order of the expansion parameter. This resummation is of course
incomplete and violates even crossing symmetry for the vertices involved in
the renormalization procedure. This causes problems concerning the
Nambu-Goldstone modes \cite{baymgrin} in the broken symmetry case or
concerning local symmetries (gauge symmetries) \cite{vHK2001} on a level
where the gauge fields are treated beyond the classical field level, i.e.,
on the propagator level.

It can be shown though, that on top of any solution of a $\Phi$-derivable
approximation which is constructed from a symmetric Lagrangian there exists
a non-perturbative effective action $\Gamma_{\text{eff}}[\varphi]$ which
generates proper vertex functions in the same sense as the 1PI effective
action. These \emph{external vertex functions} fulfill the Ward-Takahashi
identities of the underlying symmetry. However, in general they coincide
with the self-consistent ones only up to one-point order. This fact
especially ensures that the expectation values of Noether currents are
conserved for the $\Phi$-derivable approximations. Thus usually the so
generated external self-energy and higher vertex functions are different
from the $\Phi$-derivable expressions.  Details on these considerations
will be given in a forthcoming paper \cite{vHK2001-Ren-III}.

\section*{Acknowledgments}

We are grateful to J. Berges, J. P. Blaizot, G. E. Brown, P. Danielewicz,
B.  Friman, Yu. Ivanov, M. Lutz, M. A. Nowak and D.  Voskresensky for
fruitful discussions and suggestions at various stages of this work.

\begin{flushleft}

\end{flushleft}
\end{document}